\documentclass[twocolumn]{aastex63}

\usepackage{xspace}
\usepackage{enumitem}
\usepackage[utf8]{inputenc}
\usepackage{newunicodechar,graphicx}
\usepackage{amsmath}
\usepackage{lineno}
\DeclareRobustCommand{\okina}{%
  \raisebox{\dimexpr\fontcharht\font`A-\height}{%
    \scalebox{0.8}{`}%
  }%
}
\newunicodechar{ʻ}{\okina}

\newcommand{\tess}{\textit{TESS}\xspace}

\newcommand{\kepler}{\textit{Kepler}\xspace}
\newcommand{\eleanor}{\textsf{eleanor}\xspace}
\newcommand{\tesscut}{\textsf{TESSCut}\xspace}
\newcommand{\lightkurve}{\textsf{lightkurve}\xspace}
\newcommand{\exoplanet}{\textsf{exoplanet}\xspace}
\newcommand{\astropy}{\textsf{astropy}\xspace}
\newcommand{\giants}{\textsf{giants}\xspace}

\newcommand{\hoststar}{TOI-2337\xspace}
\newcommand{\hoststartwo}{TOI-4329\xspace}
\newcommand{\hoststarthree}{TOI-2669\xspace}

\newcommand{\starmass}{$1.325\pm0.118$ $M_\odot$\xspace}
\newcommand{\starradius}{$3.220\pm0.062$ $R_\odot$\xspace}
\newcommand{\teff}{$4780\pm100$ K\xspace}
\newcommand{\feonh}{$0.39\pm0.06$ dex\xspace}

\newcommand{\age}{$4.9\pm 1.8$ Gyr\xspace}
\newcommand{\starrho}{$0.039\pm0.005$ $\rho_\odot$\xspace}
\newcommand{\logg}{$3.50\pm0.06$ dex\xspace}

\newcommand{\planet}{TOI-2337b\xspace}
\newcommand{\planetmass}{$1.60\pm0.15$ $M_J$\xspace}
\newcommand{\planetradius}{$0.9\pm0.1$ $R_J$\xspace}
\newcommand{\period}{$2.99432\pm0.00008$ d\xspace}
\newcommand{\transittime}{$1793.828\pm0.003$}

\newcommand{\starmasstwo}{$1.538\pm0.049$ $M_\odot$\xspace}
\newcommand{\starradiustwo}{$2.31\pm0.03$ $R_\odot$\xspace}
\newcommand{\tefftwo}{$6000\pm100$ K\xspace}
\newcommand{\feonhtwo}{$0.29\pm0.06$ dex\xspace}

\newcommand{\agetwo}{$2.39\pm0.45$ Gyr\xspace}
\newcommand{\starrhotwo}{$0.127\pm0.006$ $\rho_\odot$\xspace}
\newcommand{\loggtwo}{$3.90\pm0.03$ dex\xspace}

\newcommand{\planettwo}{TOI-4329b\xspace}
\newcommand{\planetmasstwo}{$0.45\pm0.09$ $M_J$\xspace}
\newcommand{\planetradiustwo}{$1.50\pm0.19$ $R_J$\xspace}
\newcommand{\periodtwo}{$2.9223\pm0.00015$\xspace}
\newcommand{\transittimetwo}{$1765.068\pm0.002$}

\newcommand{\starmassthree}{$1.19\pm0.16$ $M_\odot$\xspace}
\newcommand{\starradiusthree}{$4.10\pm0.04$ $R_\odot$\xspace}
\newcommand{\teffthree}{$4800\pm100$ K\xspace}
\newcommand{\feonhthree}{$0.1\pm0.06$ dex\xspace}

\newcommand{\agethree}{$5.9\pm3.0$ Gyr\xspace}
\newcommand{\starrhothree}{$0.018\pm0.003$ $\rho_\odot$\xspace}
\newcommand{\loggthree}{$3.29\pm0.07$ dex\xspace}

\newcommand{\planetthree}{TOI-2669b\xspace}
\newcommand{\planetmassthree}{$0.61\pm0.19$ $M_J$\xspace}
\newcommand{\planetradiusthree}{$1.76\pm0.16$ $R_J$\xspace}
\newcommand{\periodthree}{$6.2034 \pm0.0001$\xspace}
\newcommand{\transittimethree}{$1521.598\pm0.009$}

\shorttitle{TESS GTG II: \planet, \planettwo, and \planetthree}
\shortauthors{Grunblatt et al.}

\graphicspath{{./}{figures/}}

\begin{document}

% \title{Discovery of a Hot Jupiter Orbiting Subgiant \hoststar}

\title{\tess Giants Transiting Giants II: The hottest Jupiters orbiting evolved stars}

\author[0000-0003-4976-9980]{Samuel K. Grunblatt}
\altaffiliation{Kalbfleisch Fellow}
\affiliation{American Museum of Natural History, 200 Central Park West, Manhattan, NY 10024, USA}
\affiliation{Center for Computational Astrophysics, Flatiron Institute, 162 5$^\text{th}$ Avenue, Manhattan, NY 10010, USA}

\author[0000-0003-2657-3889]{Nicholas Saunders}
\altaffiliation{NSF Graduate Research Fellow}
\affiliation{Institute for Astronomy, University of Hawaiʻi at M\=anoa, 2680 Woodlawn Drive, Honolulu, HI 96822, USA}

\author{Meng Sun}
\affiliation{Department of Astronomy, University of Wisconsin Madison, Madison WI 53726, USA}
% \affiliation{Department of Physics and Astronomy, Northwestern University, 2145 Sheridan Road, Evanston, IL 60208, USA}
% \affiliation{Center for Interdisciplinary Exploration and Research in Astrophysics (CIERA), 1800 Sherman Avenue, Evanston, IL 60201, USA}

\author[0000-0003-1125-2564]{Ashley Chontos}
\altaffiliation{NSF Graduate Research Fellow}
\affiliation{Institute for Astronomy, University of Hawaiʻi at M\=anoa, 2680 Woodlawn Drive, Honolulu, HI 96822, USA}

\author{Melinda Soares-Furtado}
\altaffiliation{NHFP Fellow}
\affiliation{Department of Astronomy, University of Wisconsin Madison, Madison WI 53726, USA}

\author{Nora Eisner}
\affiliation{Department of Physics, University of Oxford, Keble Road, Oxford OX1 3RH, UK}

\author{Filipe Pereira}
\affiliation{Instituto de Astrofisica e Ciencias do Espaco, Universidade do Porto, Rua das Estrelas, 4150-762 Porto, Portugal}
\affiliation{Departamento de Fisica e Astronomia, Faculdade de Ciencias da Universidade do Porto, Rua do Campo Alegre, s/n, 4169-007 Porto, Portugal}

\author[0000-0002-9258-5311]{Thaddeus Komacek}
\altaffiliation{51 Pegasi b Fellow}
\affiliation{Department of Astronomy, University of Maryland, College Park, MD 20742, USA}
\affiliation{Department of the Geophysical Sciences, The University of Chicago, Chicago, IL, 60637, USA}
% \affiliation{U. Maryland}

\author[0000-0001-8832-4488]{Daniel Huber}
\affiliation{Institute for Astronomy, University of Hawaiʻi at M\=anoa, 2680 Woodlawn Drive, Honolulu, HI 96822, USA}

\author[0000-0001-6588-9574]{Karen Collins}
\affiliation{Center for Astrophysics $\vert$ Harvard \& Smithsonian, 60 Garden St., Cambridge, MA 02138, USA}

\author[0000-0003-3092-4418]{Gavin Wang}
\affiliation{Tsinghua International School, Beijing 100084, China}

\author[0000-0003-2163-1437]{Chris Stockdale}
\affiliation{Hazelwood Observatory, Australia}

\author[0000-0002-8964-8377]{Samuel N. Quinn}
\affiliation{Center for Astrophysics $\vert$ Harvard \& Smithsonian, 60 Garden St., Cambridge, MA 02138, USA}

\author[0000-0003-1001-0707]{Rene Tronsgaard}
\affiliation{DTU Space, National Space Institute, Technical University of Denmark, Elektrovej 328, DK-2800 Kgs. Lyngby, Denmark}

\author[0000-0002-4891-3517]{George Zhou}
\affiliation{Center for Astrophysics $\vert$ Harvard \& Smithsonian, 60 Garden St., Cambridge, MA 02138, USA}
\affiliation{Centre for Astrophysics, University of Southern Queensland, Toowoomba, QLD, 4350, Australia}

\author{Grzegorz Nowak}
\affiliation{Instituto de Astrofisica de Canarias, 38205 La Laguna, Tenerife, Spain}
\affiliation{Departamento de Astrofisica, Universidad de La Laguna, 38206 La Laguna, Tenerife, Spain}

\author{Hans J. Deeg}
\affiliation{Instituto de Astrofisica de Canarias, 38205 La Laguna, Tenerife, Spain}
\affiliation{Departamento de Astrofisica, Universidad de La Laguna, 38206 La Laguna, Tenerife, Spain}

\author[0000-0002-5741-3047]{David R. Ciardi}
\affiliation{Caltech/IPAC-NASA Exoplanet Science Institute Pasadena, CA, USA}

\author{Andrew Boyle}
\affiliation{Caltech/IPAC-NASA Exoplanet Science Institute Pasadena, CA, USA}

\author[0000-0002-7670-670X]{Malena Rice}
\altaffiliation{NSF Graduate Research Fellow}
\affiliation{Department of Astronomy, Yale University, New Haven, CT 06511, USA}

\author[0000-0002-8958-0683]{Fei Dai}
\affiliation{Division of Geological and Planetary Sciences,
1200 E California Blvd, Pasadena, CA, 91125, USA}

\author{Sarah Blunt}
\affiliation{Cahill Center for Astronomy \& Astrophysics, California Institute of Technology, Pasadena, CA 91125, USA}

\author{Judah Van Zandt}
\affiliation{Department of Physics \& Astronomy, University of California Los Angeles, Los Angeles, CA 90095, USA}

\author[0000-0001-7708-2364]{Corey Beard}
\affiliation{Department of Physics \& Astronomy, University of California Irvine, Irvine, CA 92697, USA}

\author[0000-0001-8898-8284]{Joseph M. Akana Murphy}
\altaffiliation{NSF Graduate Research Fellow}
\affiliation{Department of Astronomy and Astrophysics, University of California, Santa Cruz, CA 95064, USA}

\author[0000-0002-4297-5506]{Paul A.\ Dalba}
\altaffiliation{NSF Astronomy and Astrophysics Postdoctoral Fellow}
\affiliation{Department of Astronomy and Astrophysics, University of California, Santa Cruz, CA 95064, USA}
\affiliation{Department of Earth and Planetary Sciences, University of California Riverside, 900 University Ave, Riverside, CA 92521, USA}

\author[0000-0001-8342-7736]{Jack Lubin}
\affiliation{Department of Physics \& Astronomy, University of California Irvine, Irvine, CA 92697, USA}

\author{Alex Polanski}
\affiliation{Department of Physics \& Astronomy, University of Kansas, 1082 Malott,1251 Wescoe Hall Dr., Lawrence, KS 66045, USA}

\author[0000-0002-4480-310X]{Casey Lynn Brinkman}
\altaffiliation{NSF Graduate Research Fellow}
\affiliation{Institute for Astronomy, University of Hawaiʻi at M\=anoa, 2680 Woodlawn Drive, Honolulu, HI 96822, USA}

\author{Andrew W.\ Howard}
\affiliation{Cahill Center for Astronomy \& Astrophysics, California Institute of Technology, Pasadena, CA 91125, USA}

\author{Lars A. Buchhave}
\affiliation{DTU Space, National Space Institute, Technical University of Denmark, Elektrovej 328, DK-2800 Kgs. Lyngby, Denmark}

\author[0000-0003-4540-5661]{Ruth Angus}
\affiliation{American Museum of Natural History, 200 Central Park West, Manhattan, NY 10024, USA}
\affiliation{Center for Computational Astrophysics, Flatiron Institute, 162 5$^\text{th}$ Avenue, Manhattan, NY 10010, USA}
\affiliation{Department of Astronomy, Columbia University, 550 West 120$^\text{th}$ Street, New York, NY, USA}

\author[0000-0003-2058-6662]{George R.\ Ricker}
\affiliation{Department of Physics and Kavli Institute for Astrophysics and Space Research, Massachusetts Institute of Technology, Cambridge, MA 02139, USA}

\author[0000-0002-4715-9460]{Jon M.\ Jenkins}
\affiliation{NASA Ames Research Center, Moffett Field, CA, 94035}

\author[0000-0002-5402-9613]{Bill Wohler}
\affiliation{SETI Institute, Mountain View, CA 94043}
\affiliation{NASA Ames Research Center, Moffett Field, CA, 94035}

\author{Robert~F.~Goeke}
\affiliation{Department of Physics and Kavli Institute for Astrophysics and Space Research, Massachusetts Institute of Technology, Cambridge, MA 02139, USA}

\author[0000-0001-8172-0453]{Alan~M.~Levine}
\affiliation{Department of Physics and Kavli Institute for Astrophysics and Space Research, Massachusetts Institute of Technology, Cambridge, MA 02139, USA}

\author{Knicole~D.~Colon}
\affiliation{NASA Goddard Space Flight Center, Exoplanets and Stellar Astrophysics Laboratory (Code 667), Greenbelt, MD 20771, USA}

\author[0000-0003-0918-7484]{Chelsea~ X.~Huang}
\affiliation{Department of Physics and Kavli Institute for Astrophysics and Space Research, Massachusetts Institute of Technology, Cambridge, MA 02139, USA}
\affiliation{Centre for Astrophysics, University of Southern Queensland, Toowoomba, QLD, 4350, Australia}

\author{Michelle Kunimoto}
\affiliation{Department of Physics and Kavli Institute for Astrophysics and Space Research, Massachusetts Institute of Technology, Cambridge, MA 02139, USA}

\author[0000-0002-1836-3120]{Avi~Shporer}
\affiliation{Department of Physics and Kavli Institute for Astrophysics and Space Research, Massachusetts Institute of Technology, Cambridge, MA 02139, USA}

\author[0000-0001-9911-7388]{David W. Latham}
\affiliation{Center for Astrophysics $\vert$ Harvard \& Smithsonian, 60 Garden St., Cambridge, MA 02138, USA}

\author[0000-0002-6892-6948]{Sara Seager}
\affiliation{Department of Physics and Kavli Institute for Astrophysics and Space Research, Massachusetts Institute of Technology, Cambridge, MA 02139, USA}
\affiliation{Department of Earth, Atmospheric, and Planetary Sciences, Massachusetts Institute of Technology, 77 Massachusetts Ave., Cambridge, MA 02139, USA}
\affiliation{Department of Aeronautics and Astronautics, Massachusetts Institute of Technology, 77 Massachusetts Ave., Cambridge, MA 02139, USA}

\author{Roland K.\ Vanderspek}
\affiliation{Department of Physics and Kavli Institute for Astrophysics and Space Research, Massachusetts Institute of Technology, Cambridge, MA 02139, USA}

\author[0000-0002-4265-047X]{Joshua N.\ Winn}
\affiliation{Department of Astrophysical Sciences, Princeton University, 4 Ivy Lane, Princeton, NJ 08544, USA}

% \author{Rafael Brahm}
% \affiliation{Center of Astro-Engineering UC Pontificia Universidad Cat\'olica de Chile, Av. Vicu\~na Mackenna 4860, 7820436 Macul, Santiago, Chile}
% \affiliation{Instituto de Astrof\'isica Pontificia Universidad Cat\'olica de Chile, Av. Vicu\~na Mackenna 4860, 7820436 Macul, Santiago, Chile}
% \affiliation{Millennium Institute for Astrophysics, Chile}

% \author{Andr\'es Jord\'an}
% \affiliation{Millennium Institute for Astrophysics, Chile}
% \affiliation{Facultad de Ingenier\'ia y Ciencias, Universidad Adolfo Ib\'a\~nez, Av. Diagonal las Torres 2640, Pe\~nalol\'en, Santiago, Chile}

% \author{Sam Quinn}
% \affiliation{Center for Astrophysics $\vert$ Harvard \& Smithsonian, 60 Garden St., Cambridge, MA 02138, USA}

% \author{Andrew Vanderburg}
% \author{Ruth Angus}
% \author{So Hattori}

% \author{additional collaborators}
% \affiliation{still need to add these}

\begin{abstract}

Giant planets on short-period orbits are predicted to be inflated and eventually engulfed by their host stars. However, the detailed timescales and stages of these processes are not well known. Here we present the discovery of three hot Jupiters (P $<$ 10 d) orbiting evolved, intermediate-mass stars ($M_\star$ $\approx$ 1.5 M$_\odot$, 2 R$_\odot$ $<$ $R_\star < $ 5 R$_\odot$). By combining \tess photometry with ground-based photometry and radial velocity measurements, we report masses and radii for these three planets between 0.4 and 1.8 M$_\mathrm{J}$ and 0.8 and 1.8 R$_\mathrm{J}$. \planet has the shortest period (P=\period) of any planet discovered around a red giant star to date. Both \planettwo and \planetthree appear to be inflated, but \planet does not show any sign of inflation. The large radii and relatively low masses of \planettwo and \planetthree place them among the lowest density hot Jupiters currently known, while \planet is conversely one of the highest. All three planets have orbital eccentricities below 0.2. The large spread in radii for these systems implies that planet inflation has a complex dependence on planet mass, radius, incident flux, and orbital properties. We predict that \planet has the shortest orbital decay timescale of any planet currently known, but do not detect any orbital decay in this system. Transmission spectroscopy of \planettwo  would provide a favorable opportunity for the detection of water, carbon dioxide and carbon monoxide features in the atmosphere of a planet orbiting an evolved star, and could yield new information about planet formation and atmospheric evolution.

%Transmission spectroscopy of \planettwo would provide the best chances to detect water, carbon dioxide and carbon monoxide features in a planetary atmosphere around an evolved star, tracers of planetary formation and atmospheric evolution. %Tracing planetary evolution from precursor younger planets identified by other surveys to these and other evolved systems will constrain planet inflation mechanisms and other star-planet interactions.

%Additionally, the high metallicities of these evolved host stars may be evidence for atmospheric stripping. 

    % Further observations with \tess may allow detection of asteroseismic oscillations, key to determining more accurate parameters for this system.

    % We present the discovery and validation of a planet around \hoststar (\tic). While the population of confirmed exoplanets continues to grow, the sample of confirmed planets around evolved stars is limited, and much is still unknown about these systems. \hoststar is a roughly solar mass star ($M_*=1.02\pm 0.120309 M_\odot$) that has started evolving off the main sequence, expanding to a radius of $R_*=3.15138\pm 0.169143 R_\odot$. \planet went unidentified by existing planet searches due to periodic \tess systematics. Using combined \tess photometry, ground-based photometry, and ground-based radial velocity measurements, we report a planet radius of $R_p$=\planetradius and planet mass of $M_p=$\planetmass. Additionally, we discuss the measures taken to improve planet detection capabilities for FFIs and the potential subsets of planets that remain undiscovered.

\end{abstract}

\section{Introduction} \label{sec:intro}

%planets arnd evolved stars
%Discoveries over the last 30 years have shown us that our solar system is not unique. Fossil records give us {\it in situ} samples of the past of our planet, but to understand what the future will hold, we must rely on observations of exoplanetary systems more evolved than our own. 

Exoplanets have been known around evolved stars at orbital separations $>$1 AU for decades \citep[e.g.][]{hatzes2003}. However, planets at smaller separations were expected to be engulfed due to angular momentum exchange through tides \citep{hut1981,villaver2009}. Thus less than a decade ago, no planets were known on orbits smaller than 1 AU around stars with radii greater than 3 R$_\odot$ and masses above 1 M$_\odot$ \citep{schlaufman2013, villaver2014}.

This changed with the discovery of Kepler-91b, a Jupiter-sized planet orbiting a 6.3 R$_\odot$ star every 6.25 days \citep{lillobox2014,barclay2015}. Subsequent discoveries with the NASA \emph{Kepler} Mission and its extension, \emph{K2}, have proven planets can survive at short periods around evolved stars \citep{almenara2015, vaneylen2016,chontos2019}. Detailed stellar characterization through asteroseismology of these systems revealed that planets can interact with their host star, potentially becoming re-inflated at late times \citep{grunblatt2016,grunblatt2017}. Occurrence studies have now revealed that hot Jupiters are equally common around main sequence stars and low-luminosity red giant branch stars \citep{howard2012, grunblatt2019}. However, a number of questions about the origins of these systems remain unanswered: Were these planets hot Jupiters before their host stars evolved off the main sequence, or cold Jupiters that have migrated inwards thanks to tidal interactions with their post-main sequence host stars \citep{grunblatt2018}? Are the atmospheres of these planets undergoing stripping \citep[e.g.,][]{spake2018, bell2019}? Other observable properties of red giants are said to be potential signs of planet engulfment \citep{carlberg2013,macleod2018,soaresfurtado2020}. Do these correlate strongly with hot Jupiter occurrence? Can we detect clear evidence for orbital decay \citep{yee2020,turner2021}, planetary atmospheric evolution \citep{baxter2021}, or exchange of material between star and planet \citep[e.g.,][]{Rappaport2013}, to constrain the death of a planetary system?

%how great is tess right? 

The Transiting Exoplanet Survey Satellite (\tess; \citealt{ricker2014}) is enabling the discovery of a predicted $\sim$14,000 planets \citep{sullivan2015,barclay2018}. During its nominal 2-year mission, the space telescope observed most of its targets in the Full Frame Images (FFIs) with a 30-minute observing cadence, and has completed at least one year of observations in each of the northern and southern hemispheres. Each year was split into 13 observing sectors that stretched from the ecliptic pole to the ecliptic plane, moving every $\sim$27 days. Targets near the ecliptic pole were observed in multiple sectors, in some cases providing a full year of photometry, while targets closer to the ecliptic plane were observed in fewer sectors. According to the NASA Exoplanet Science Institute (NExSci) archive\footnote{\href{https://nexsci.caltech.edu}{nexsci.caltech.edu}}, \tess has already led to the discovery of 100+ confirmed planets and 4,000+ project candidates \citep{guerrero2021}. Of the planets confirmed to date, only a handful orbit evolved host stars; among those is TOI-197b, the first \tess planet discovery orbiting an evolved host with an asteroseismic detection \citep{huber2019}. There have also been detections of planets orbiting subgiant stars for which asteroseismic detections were not possible \citep[e.g.,][]{nielsen2019,wang2019,eisner2020,saunders2021}.

Through our programs to identify and confirm planets around evolved stars using \tess FFIs (Guest Investigator programs GO22102,GO3151,GO4179), we have begun successfully identifying new planet candidates and confirming new planetary systems. The first planet found in our survey was observed in both 30-minute and 2-minute cadence \tess\ data \citep{saunders2021}. Here we present three additional planets confirmed by our survey, all found in 30-minute cadence, full frame image \tess\ data. These planets are among the shortest period planets found around evolved host stars, and thus provide new constraints on the process of planetary inspiral and engulfment. Here we present analysis of the currently available data, and discuss what future observations from \tess\ and other ground-based and space-based facilities of these systems may reveal about star-planet interaction and evolution.

%what remains unknown about this population? what have we figured out so far?

%what did we find?

%This paper is laid out as follows: \S \ref{sec:methods} describes the observations and data analysis applied to detect the three planets, \S \ref{sec:hoststar} and \ref{sec:planetprops} lay out the properties of the systems, \S \ref{sec:discussion} discusses our results in the context of planet populations, and \S \ref{sec:conclusions} lays out our conclusions.

\section{Observations} \label{sec:methods}

\subsection{\tess Photometry} \label{sec:photo}

\begin{figure*}[ht!]
    \centering
    \includegraphics[width=0.95\textwidth]{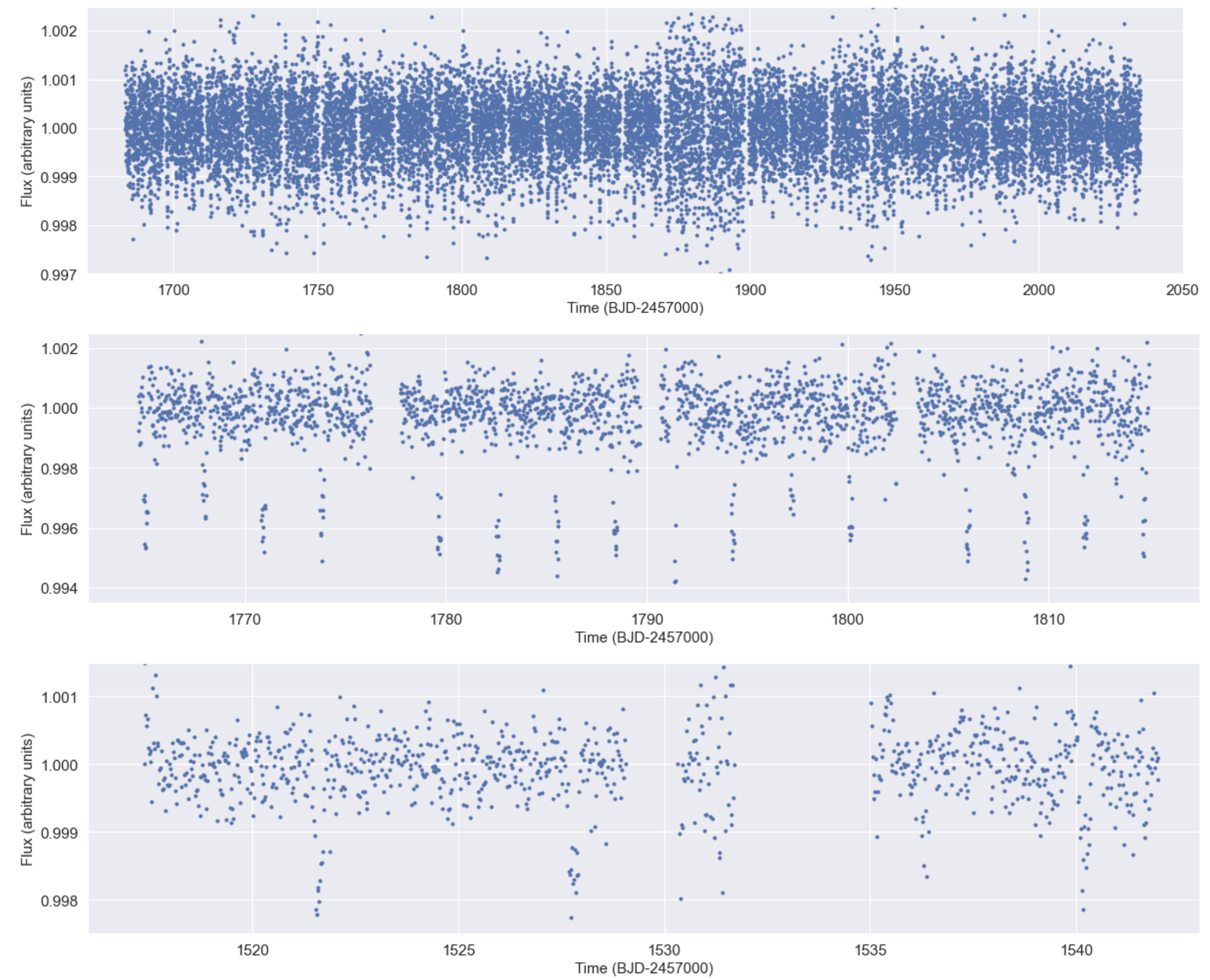}
    \caption{Full \tess\ prime mission light curves of \hoststar, \hoststartwo, and \hoststarthree (from top to bottom, respectively). Light curves for all stars in this study were produced through the \texttt{giants} pipeline. Additional data have since been acquired at 10-minute cadence for \hoststar and \hoststarthree, supporting the previous transit detections.}
    \label{fig:unfolded_lc}
\end{figure*}

\planet, \planettwo and \planetthree were discovered as part of our survey identifying new planets around evolved host stars. Using the \tess Input Catalog \citep[TICv8;][]{stassun2019}, we made cuts based on color, magnitude, and Gaia parallax in order to limit our sample to evolved stars. We developed the \giants\footnote{\href{https://github.com/nksaunders/giants}{https://github.com/nksaunders/giants}} Python package for accessing, de-trending, and searching \tess observations for periodic transit signals \citep{saunders2021}. The details of how this pipeline processes \tess\ full frame image data are described in detail in \citet{saunders2021}. We present the \giants\ light curves for these targets in Figure \ref{fig:unfolded_lc}.

% The \giants pipeline uses the \tesscut tool to access \tess\ FFI data calibrated by the \tess\ Science Processing Operations Center \citep{jenkins2016,brasseur2019}. The scattered Earth light is then removed by creating a design matrix $F$ from the flux light curves of each pixel outside the target aperture mask in a 11x11 cutout. This target aperture mask was chosen using an optimized thresholding method, which was tuned by hand for the three planet-hosting systems analyzed here. The optimized noise model was subtracted from the raw flux light curve to produce a background-corrected light curve \citep{saunders2021}. This procedure is similar to the Pixel Level Decorrelation method applied to the \textit{Spitzer} Space Telescope by \cite{deming2015} and the \textit{K2} mission by \cite{luger2016,luger2018}. Figure \ref{fig:unfolded_lc} shows the full length light curves made for each of these targets. 

% In addition, the \giants pipeline produces a summary PDF for each target. The \giants pipeline summary includes the full detrended light curve, a Lomb-Scargle \citep{lomb1976, scargle1982} periodogram to search for stellar oscillations as well as a box-least squares (BLS) periodogram to search for transiting planets. The flattened light curve is folded to the period corresponding to the maximum power from the periodogram, which includes a basic transit fit using the Python package ktransit \citep{barclay2015}. Finally, the folded light curve is further separated into even and odd transits to identify any depth differences indicative of false positive scenarios.

We used our \texttt{giants} pipeline to produce \tess\ light curves for as many red giant branch stars with \tess\ magnitude $m_T <$ 13 as possible. We produced approximately 540,000 light curves from the first 2 years of data from the \tess\ Mission. We then performed an automated BLS search on these targets, and produced summary plots using the BLS output as well as TIC information and the pixel cut out. These summary plots were then visually inspected, during which all three of these candidates were flagged for potential rapid followup.

13 sectors of data were available for \planet at the conclusion of the \tess\ prime mission, during which over 100 transits were observed. Additional data is currently being taken for this target as part of the \tess\ extended mission. Conversely, \planettwo was only observed in three sectors of the \tess\ prime mission. The last sector of data for \planettwo also appeared to be corrupted when producing our \texttt{giants} light curve, and thus only 15 transits were observed. \planetthree was only observed for one sector of the \tess\ prime mission, during which three planet transits were observed. 4 additional planet transits were observed during one additional sector of observation in the \tess\ extended mission, which we include in our full planet characterization in Section \ref{sec:planetprops}.

%produces a one-page PDF summary for each target including the full de-trended light curve (de-trending methods described in detail below), a Lomb-Scargle periodgram \citep{lomb1976, scargle1982} to identify stellar oscillations in the flux light curve, the box least squares (BLS) periodogram, a flattened light curve folded with the period of maximum power in the BLS periodogram, as well as folded light curves of exclusively even and odd transits to identify the existence of a depth difference, and finally a transit fit using the \textsf{ktransit} Python package \citep{barclay2015}.

To ensure our transit depths from the \giants pipeline were reliable, we also generated a \tess light curve for each of \hoststar, \hoststartwo and \hoststarthree using the \eleanor pipeline to access FFI data calibrated by the \tess\ Science Processing Operations Center \citep{jenkins2016,feinstein2019}, and performed our transit search on the \eleanor-corrected light curves. We additionally performed our own systematics removal on the \eleanor-corrected light curves through a modified version of the \giants pipeline to independently verify the presence of transits \citep{saunders2021}. When we applied the same BLS search to our corrected light curve, we identified an eclipse signal with the same period as found in the \eleanor light curves, and transit depths that agreed within $\approx$ 10\% or less for \planet and \planetthree. 

\begin{figure}[h!]
    \centering
    \includegraphics[width=0.495\textwidth]{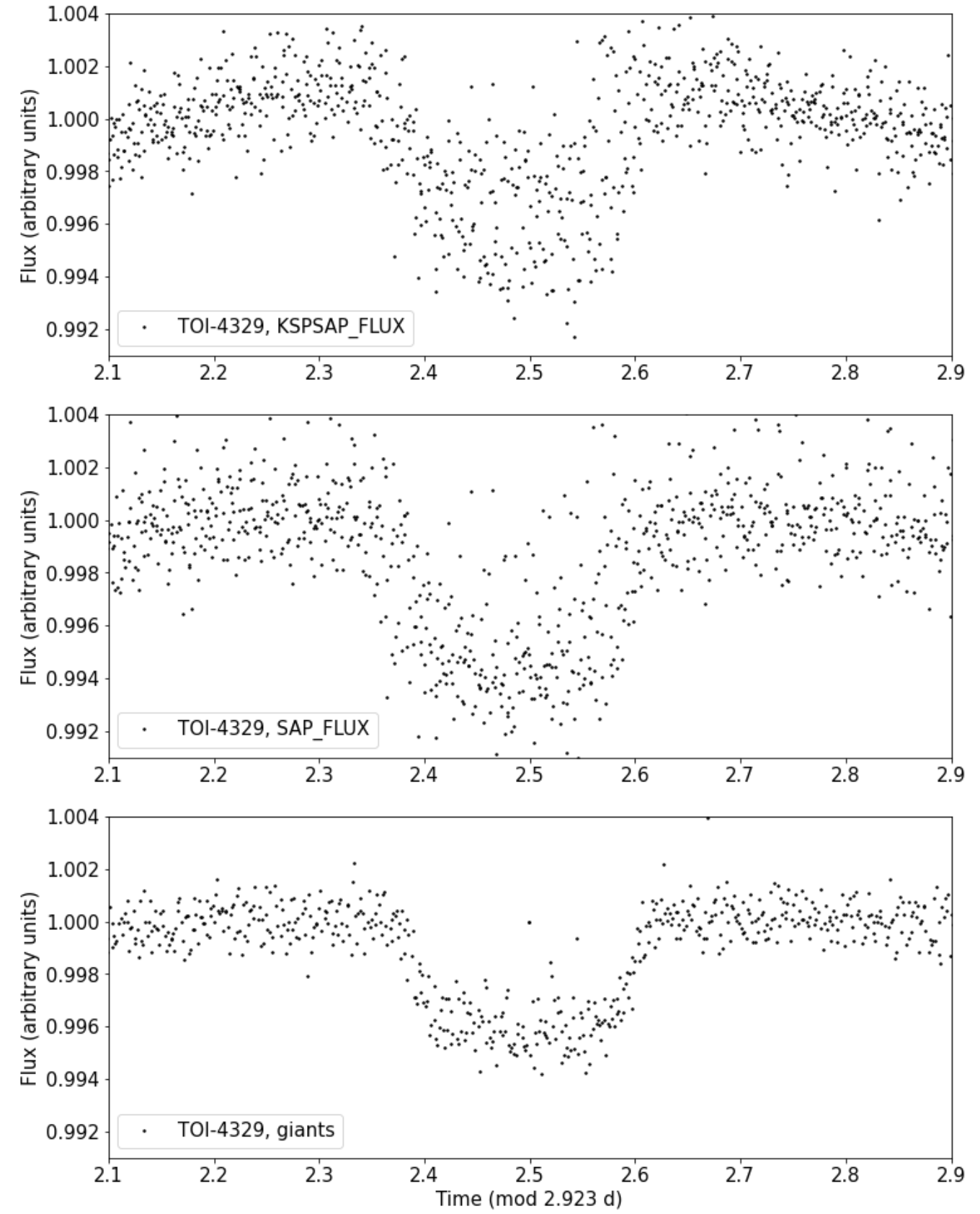}
    \caption{Phase-folded light curves of \hoststartwo using the QLP `KSPSAP FLUX', `SAP FLUX', and \texttt{giants} fluxes (from top to bottom, respectively). Clear differences in transit depth and shape and light curve scatter can be seen between the different light curves.}
    \label{fig:2567_comp}
\end{figure}

Due to a nearby star with a $\Delta m_V$ = 3.5 falling onto the same \tess\ pixel as \planettwo, flux contamination is a larger concern for this system than for \planet or \planetthree, resulting in transit depths between pipelines that disagree by $>$30\%. In order to understand the effect of flux contamination on our estimated transit depth, we compared our \texttt{giants} light curve to the QLP light curve produced by the MIT team. We display the phase-folded light curves from \texttt{giants} along with the QLP light curve, illustrating both the `KSPSAP FLUX' and `SAP FLUX' values produced by their pipeline in Figure \ref{fig:2567_comp}. As the `KSPSAP FLUX' pipeline is smoothed using a window length of 0.3 days, it is likely that the transit of \planettwo could have been diluted. This hypothesis is supported by the ramping before and after transit that is only seen in the `KSPSAP FLUX' light curve but not in the `SAP FLUX' and \texttt{giants} light curves. We illustrate the differences between these light curves in Figure \ref{fig:2567_comp}, and discuss the implications for our determination of the radius of \planettwo in Section \ref{sec:planetprops}.

\subsection{Radial Velocity Measurements}

RV observations were taken with the HIRES spectrograph on the Keck-I telescope on Maunakea, Hawaii \citep{vogt1994}. HIRES has a resolving power of $R\approx120,000$ and wavelength coverage between $\sim350$nm and $\sim620$nm. 7 RV measurements were taken of \hoststar between August 1, 2020 and September 2, 2020, 9 RV measurements were taken of \hoststartwo between August 7, 2020 and December 25, 2020, and 10 RV measurements of \hoststarthree were taken between December 31, 2020 and June 13, 2021. 

Additional RV observations of \hoststar were taken with the FIES spectrograph onboard the Nordic Optical Telescope on La Palma, Canary Islands \citep{telting2014}. FIES has a resolving power of $R\approx65,000$ and wavelength coverage between $\sim370$nm and $\sim730$nm. 14 observations with FIES were made between September 2, 2020 and and October 25, 2020.

% Additional observations of \hoststarthree were taken using the CHIRON spectrograph on board the 1.5m SMARTS telescope in Chile \citep{tokovinin2013} between November 2020 and May 2021. Data were obtained in slicer mode, which uses an image  slicer and fiber bundle to yield R $\approx$ 79,000 over the spectral range 410 nm to 880 nm. We extracted RVs by modeling the least-squares deconvolution spectral line profiles \citep{donati1997}. We also extracted RVs from this dataset through a self cross-correlation method and found results which strongly agreed with the least-squares deconvolution results presented here.

Additional observations of \hoststarthree were taken using the CHIRON spectrograph on board the 1.5m SMARTS telescope in Chile \citep{tokovinin2013} between November 2020 and May 2021. Data were obtained in slicer mode, which uses a fiber feed and an image slicer to yield R $\approx$ 79,000 over the spectral range 410 nm to 880 nm. The spectra were extracted as per \citet{paredes2021}. Radial velocities were measured from the spectra via a cross correlation between each epoch against a median-combined observed template. Approximate absolute velocities were first derived from each spectrum as per \citet{zhou2020}. The spectra were then median combined with the approximate velocities removed to form a master template spectrum. Cross correlations were performed over the wavelength range of 410 nm to 620 nm, with the Balmer, Sodium, and telluric regions masked out.

\subsection{Ground-based Imaging of \hoststar} 

\begin{figure}[ht!]
    \centering
    \includegraphics[width=.45\textwidth]{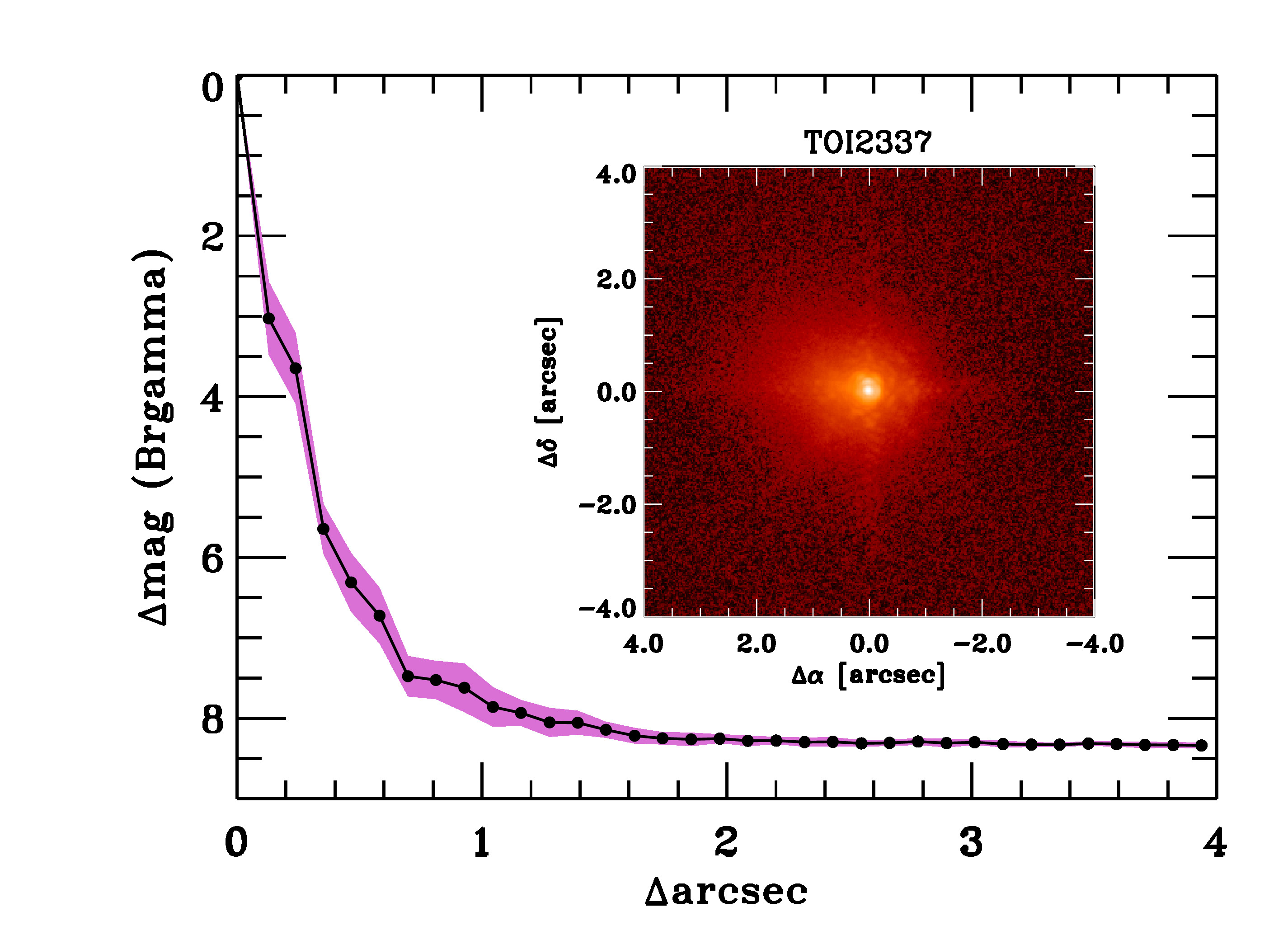}
    \caption{Contrast curve for \hoststar, with the corresponding ground-based adaptive optics image shown in the inset. \hoststar appears to be a single star based on this analysis. }
    \label{fig:imaging}
\end{figure}

\hoststar was also observed by the PHARO instrument on the 5m Hale telescope at the Palomar Observatory in California. Observations were taken on June 24, 2021 in the Br$\gamma$ filter near 2.2 $\mu$m. The contrast curve for these observations can be seen in Figure \ref{fig:imaging}, which shows the detection limits in contrast ($\Delta$m) versus angular separation from the point spread function center in arcseconds for the filter wavelength.  The inset image is the speckle auto-correlation function for the observation. No additional stars can be identified within 4'' of our target, as shown in the inset image. Although another faint star can be identified approximately 7'' southeast of our target, this star has an inconsistent proper motion with our target \citep{gaia2018} and thus is likely not associated with our target, and is too faint to cause significant transit dilution.

While \hoststartwo and \hoststarthree do not have high resolution imaging results, the Gaia astrometric noise metric RUWE is low (0.91 and 1.04, respectively), indicative that the stars are not in wide binary systems which could be resolved by \emph{Gaia} photometry. Additionally, no evidence of a spectroscopic binary can be seen in the spectra of these stars, placing limits on close binarity for these stars. The radial velocity measurements of these systems do not show any significant linear or quadratic trends with time, suggesting these stars are single and not part of a binary system.

\subsection{Ground-based Photometry of \hoststar} 

\hoststar was also observed by the Las Cumbres Observatory Global Telescope (LCOGT) 1-m telescope at McDonald Observatory in West Texas on June 12, 2021, and August 22, 2021, and by the LCOGT 1-m telescope at Teide Observatory in the Canary Islands between August 1, 2021 and August 22, 2021.

Observations from the Teide Observatory captured multiple ingresses of transits, while observations from McDonald observatory were able to capture the egresses of multiple transits. Observations were scheduled using the {\sf TESS Transit Finder}, which is a customized version of the {\sf Tapir} software package \citep{Jensen:2013}. The images were calibrated by the standard LCOGT {\sf BANZAI} pipeline \citep{McCully:2018}, and photometric data were extracted with {\sf AstroImageJ} \citep{Collins:2017}. A fit to the ground-based data using the \tess\-determined ephemeris finds a fractional transit depth $\delta \approx 800 {\rm ppm}$, orbital period ${\rm P}=2.9943209\pm0.0000387~{\rm days}$  (using the reference epoch we derive from the TESS data), and time of transit T$_0$ = 2459033.36898 $\pm$ 0.00267, within $1\sigma$ of values extracted from our simultaneous fit to the \tess \texttt{giants} light curve and RV data (see Table \ref{table:planet}). The observed ingresses and egresses of the planet transit are within 0.05 d of the expected times, corresponding to a transit ephemeris uncertainty $\Delta$T$_c$ = 40 min, placing constraints on the rate of orbital decay of this system which are discussed in more detail in Section \ref{sec:orbdec}.

\section{Host Star Characterization} \label{sec:hoststar}

%% eisner+2020
% TOI813b is in orbit around a subgiant host. The subgiant phase in some respects mimics the pre-main sequence phase in reverse: the same star-planet interaction mechanisms – tides, orbital migration, stellar heating and photoevapora- tion – are at play, but their effect is increasing, rather than decreasing, with time. Using a MIST Version-1.2 stellar evo- lution track (Choi et al. 2016), with input stellar parameters as given in Table 1 and a rotation velocity of 0 ms−1, we estimate the main sequence lifetime of TOI 813 to be ∼3.45 Gyrs. This means that, at the current age of 3.73 ± 0.62 Gyr, the target has only recently left the main sequence and is 
%%

\begin{figure}[ht!]
    \centering
    \includegraphics[width=.5\textwidth]{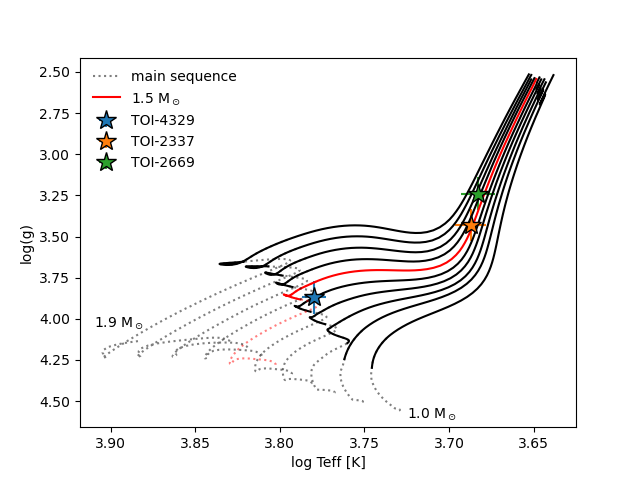}
    \caption{Position of \hoststar, \hoststartwo and  \hoststarthree on an H-R diagram. All host stars have evolved off of the main sequence onto the subgiant and red giant branch. We also illustrate MIST evolutionary tracks of 1-2 M$_\odot$, +0.25 [Fe/H] dex stars in 0.1 M$_\odot$ increments for reference. We have highlighted a MIST evolutionary track for a 1.5 M$_\odot$, [Fe/H] = 0.25 dex star in red, illustrating the evolutionary sequence probed here.}
    \label{fig:hr}
\end{figure}

\begin{figure}[ht!]
    \centering
    \includegraphics[width=.45\textwidth]{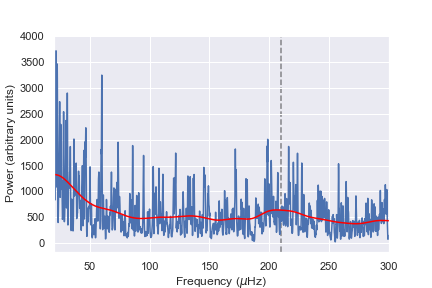}
    \caption{Power density spectrum of the 10-minute cadence data of \hoststarthree, where the data are shown in blue, a smoothed power spectrum is shown in red, and the expected maximum power of oscillation $\nu_\mathrm{max}$ is labeled by the gray dashed line. A power excess is slightly significant in the data near this value as measured with \texttt{pySYD}.}
    \label{fig:astero_3488}
\end{figure}

\subsection{High-Resolution Spectroscopy}

We used SpecMatch to measure the metallicity, surface gravity and effective temperature of the host stars from our HIRES spectra \citep{petigura2015}. We then used \textsf{isoclassify} \citep{huber2017} to combine TICv8 and spectroscopic information to determine stellar properties, listed in Table \ref{table:stellar}.

%hoststar has begun evolving off of the main sequence and onto the subgiant branch, increasing in radius and luminosity while its temperature drops. We estimate the mass of \hoststar to be \starmass by comparing SpecMatch parameters to stellar evolutionary tracks. 

Figure \ref{fig:hr} shows an H-R diagram with evolutionary tracks downloaded from the MESA Isochrones \& Stellar Tracks (MIST; \citealt{dotter2016}; \citealt{choi2016}; \citealt{paxton2011}). As all three host stars have roughly the same mass and metallicity (M$_*$ $\approx$ 1.5 M$_\odot$, [Fe/H] $\approx$ 0.25 dex), we suggest that these systems may represent an evolutionary sequence for post-main sequence, intermediate-mass stars. We find that both \hoststar and \hoststarthree lie on the red giant branch, while \hoststartwo lies at an earlier subgiant stage of evolution.

\subsection{Asteroseismology of \hoststarthree}

Asteroseismology is the study of oscillations in stars. In particular, solar-like oscillations can be used to probe the internal structure and fundamental physical properties of a star \citep{kjeldsen1995}. These oscillations can be seen and measured in the power spectrum of a light curve produced from 30-minute cadence full frame image \tess\ data of a red giant star \citep{silvaaguirre2020,grunblatt2021}. Figure \ref{fig:astero_3488} illustrates the power spectrum of the second sector of \tess\ data for \hoststarthree, where the expected frequency of excess power has been labeled by a gray dashed line. To perform a rigorous asteroseismic analysis, we use the pySYD package built on the SYD asteroseismic pipeline \citep{huber2009,chontos2021}. We identify a $\nu_\mathrm{max}$ value of 210 $\pm$ 30 $\mu$Hz for this star, corresponding to a log($g$)$\approx$3.2, in agreement with the TICv8 and \texttt{isoclassify}-determined values. However, the signal-to-noise of the detection is currently too low to determine an accurate $\Delta\nu$ value. Thus we validate our stellar mass and radius using these asteroseismic parameters, but use the \texttt{isoclassify}-derived stellar parameters to determine planet mass and radii for \planetthree. Additional asteroseismic studies with \tess\ will improve methods for identifying and measuring asteroseismic signals with limited time series data \citep[e.g.,][]{mackereth2021, hon2021}.

%Figure \ref{fig:astero_3488} illustrates the power spectrum of the second sector of \tess\ data for \hoststarthree. The left panel illustrates the unfolded power spectrum, where the expected frequency of excess power has been labeled by a gray dashed line. The right panel shows the folded power spectrum, folded to match the regular frequency spacing of the star, $\Delta\nu$. To perform a rigorous asteroseismic analysis, we use the pySYD package built on the SYD asteroseismic pipeline \citep{chontos2021, huber2009}. We identify a $\nu_\mathrm{max}$ value of 210 $\pm$ 30 $\mu$Hz for this star. However, the signal-to-noise of the detection is too low to determine an accurate $\Delta\nu$ value. Thus we validate our stellar mass and radius using these asteroseismic parameters, but use the \tess\ Input Catalog parameters to determine planet mass and radii for \planetthree.

\begin{table*}
\centering
    \begin{tabular}{l c c c}
        \hline
        \rule{0pt}{3ex}\textit{Target IDs} & & & \\
        \rule{0pt}{3ex}TOI & 2337 & 4329 & 2669\\
        TIC & 230001847 & 256722647 & 348835438 \\
        TYC & 4217-01423-1 & & 5456-00076-1 \\
        2MASS & J19222878+6051140 & J21581534+7107538 & J08585340-1318450 \\
        Gaia DR2 & 2239684947894267392 & 2224458777730383616 & 5735664051960550144\\
        \hline
        \rule{0pt}{3ex}\textit{Coordinates} & \\
        \rule{0pt}{3ex}RA(J2015.5) & 19:22:28.77 & 21:58:15.34 & 08:58:53.43\\
        Dec(J2015.5) & 60:51:14.05 & 71:07:53.78 & -13:18:45.29\\
        \hline
        \rule{0pt}{3ex}\textit{Characteristics} & \\
        \rule{0pt}{3ex}\tess\ mag & 11.23 & 11.88 & 9.99\\
        Radius $R_\star$ $(R_\odot)$ & \starradius & \starradiustwo & \starradiusthree \\
        Mass $M_\star$ $(M_\odot)$ & \starmass & \starmasstwo & \starmassthree\\
        $T_{\rm eff}$ (K) $ $ & \teff & \tefftwo & \teffthree \\
        $\log(g)$ (dex) & \logg & \loggtwo & \loggthree \\
        $ $[Fe/H] (dex) $ $ & \feonh & \feonhtwo & \feonhthree \\
        Age (Gyr) $ $ & \age & \agetwo & \agethree \\
        Density $\rho_\star$ $(\rho_\odot)$ & \starrho & \starrhotwo & \starrhothree \\
        \hline
   \end{tabular}
	 \caption{Stellar properties derived from an \textsf{isoclassify} fit to HIRES spectroscopic observations.}
	 \label{table:stellar}
\end{table*}

% \subsection{Stellar Rotation}

% An analysis of stellar variability using data collected by the Wide Angle Search for Planets (WASP) South found no rotational modulation in the range from 2 to 100 days. This was conducted using 24,000 data points from four consecutive years, covering a span of about 160 nights each year. The upper limit of photometric variability detection for this target made by WASP-South is roughly 0.8 millimagnitudes. 

\section{Planet Characterization} \label{sec:planetprops}

\subsection{Model Fit}

We used the \exoplanet Python package to simultaneously fit a model to the photometry and radial velocity observations \citep{exoplanet:exoplanet}. The data input to our model were all radial velocity observations and all sectors of \tess FFI photometry available from the first three years of the \tess\ Mission. Our model used stellar parameters derived using \textsf{isoclassify} \citep{huber2017} with \emph{Gaia} parallax, and input effective temperature ($T_\mathrm{eff}$) and metallicity estimated from spectral observations taken by the Keck observatory. These model input parameters can be found in Table \ref{table:stellar}. 

Our initial choices of planet period and depth were taken from the BLS search determined values produced during the transit search described in \S 2.2. For limb darkening, we use the quadratic model prescribed by \cite{exoplanet:kipping13} to provide a two-parameter model with uninformative sampling. We parameterized eccentricity using the single planet eccentricity distribution of \citet{vaneylen2019}. 

\begin{figure*}[ht!]
    \centering
    \includegraphics[width=.47\textwidth]{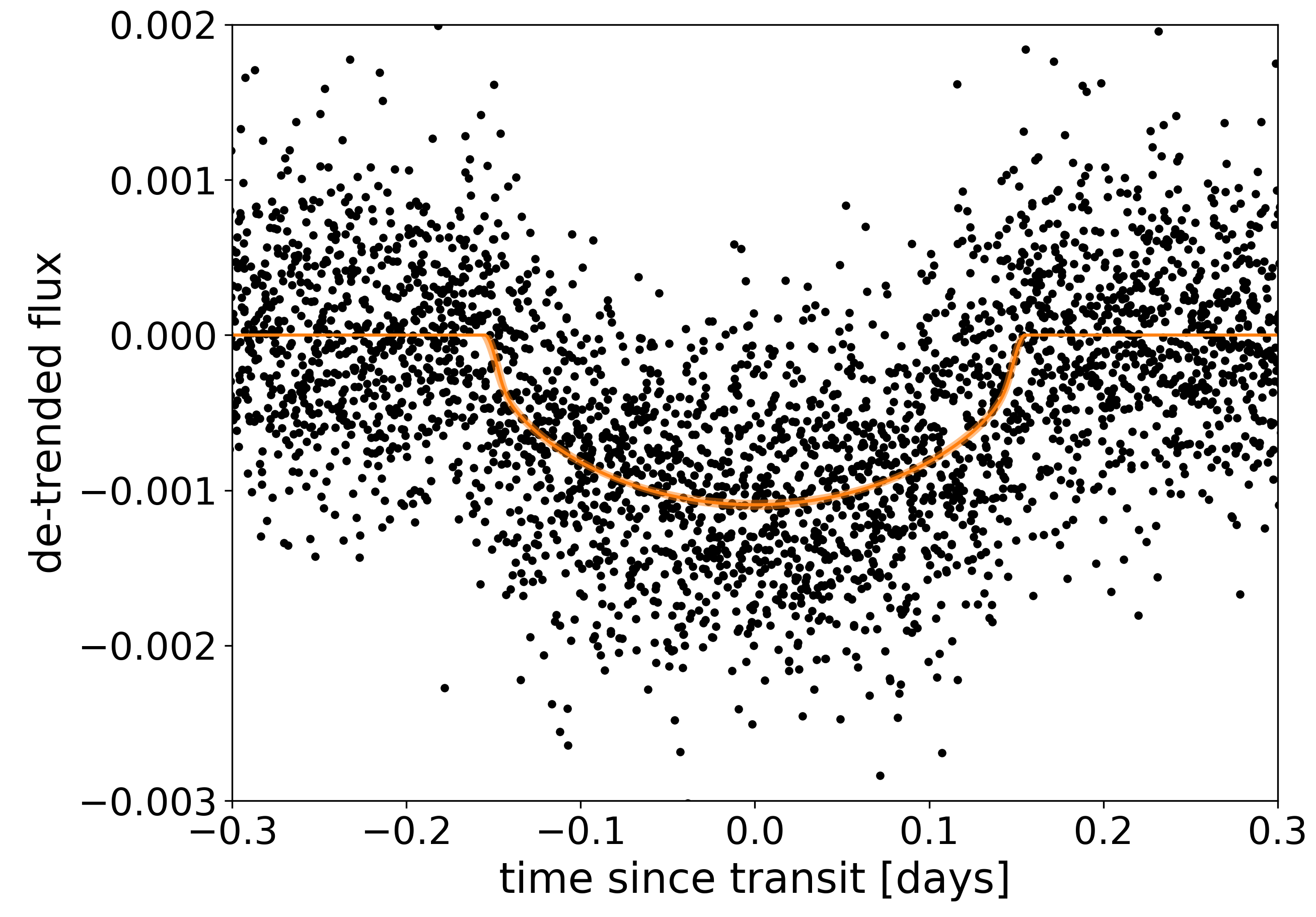}
    \includegraphics[width=.47\textwidth]{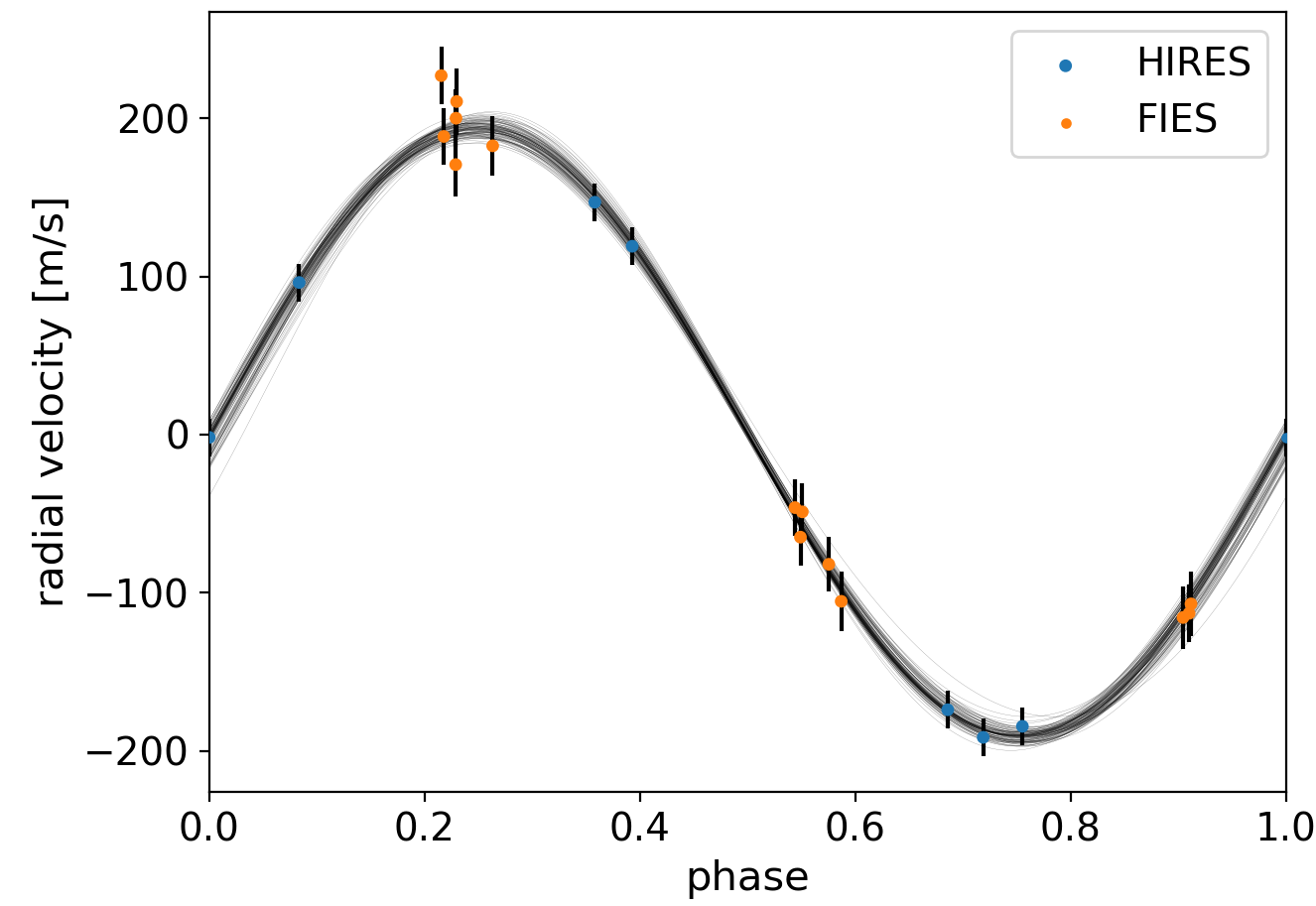}
    \caption{{\it Left:} The flux light curve of \hoststar folded at a period of \period days. The de-trended photometry is shown in black with the binned photometry overplotted in orange. {\it Right:} All radial velocity observations of \planet used in this analysis where the time axis has been folded at the orbital period of the planet. Observations come from the Nordic Optical Telescope in La Palma (orange) and the Keck-I telescope on Maunakea (blue).}
    \label{fig:lc_rv_2300}
\end{figure*}

\begin{figure*}[ht!]
    \centering
    \includegraphics[width=.5\textwidth]{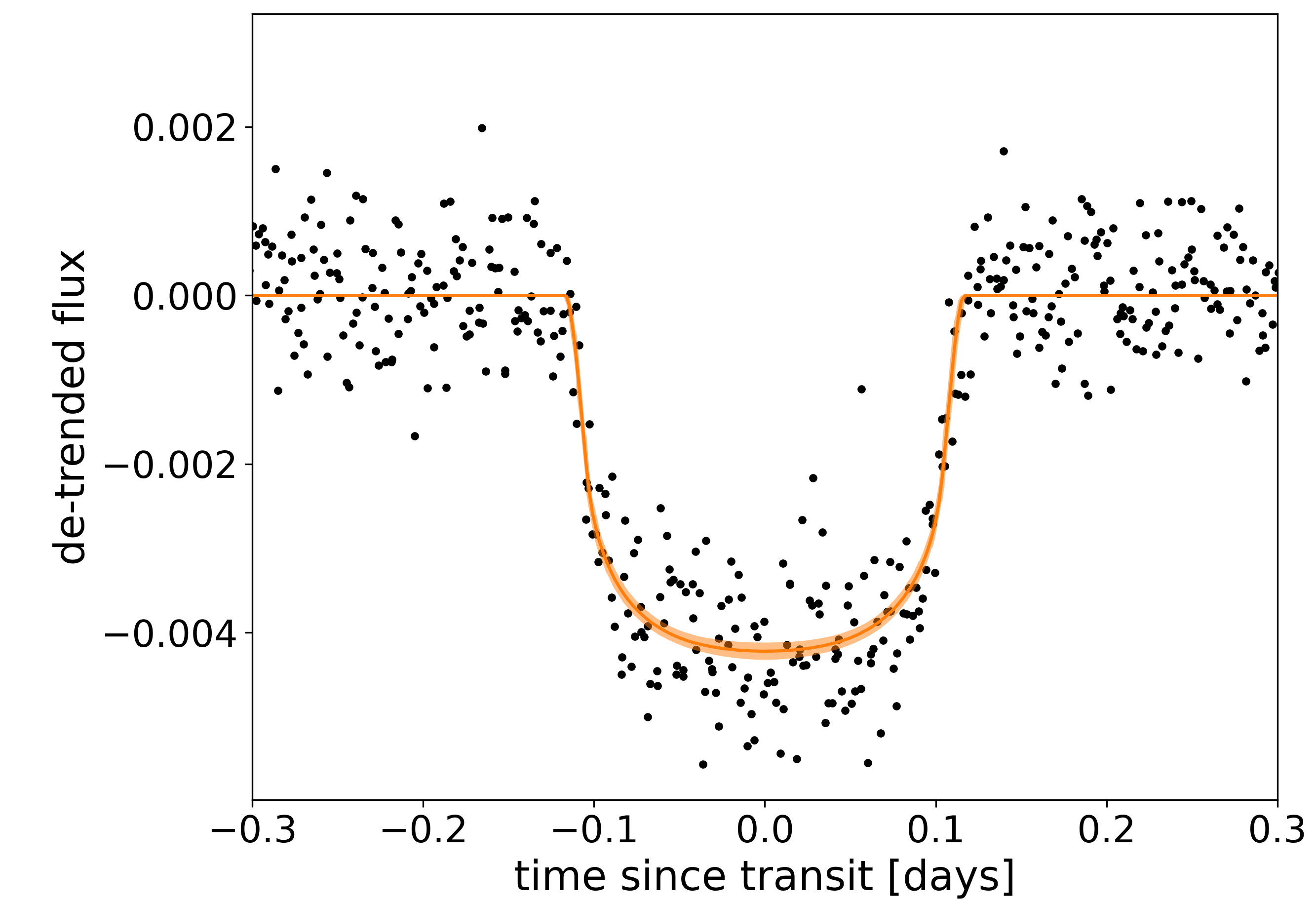}
    \includegraphics[width=.48\textwidth]{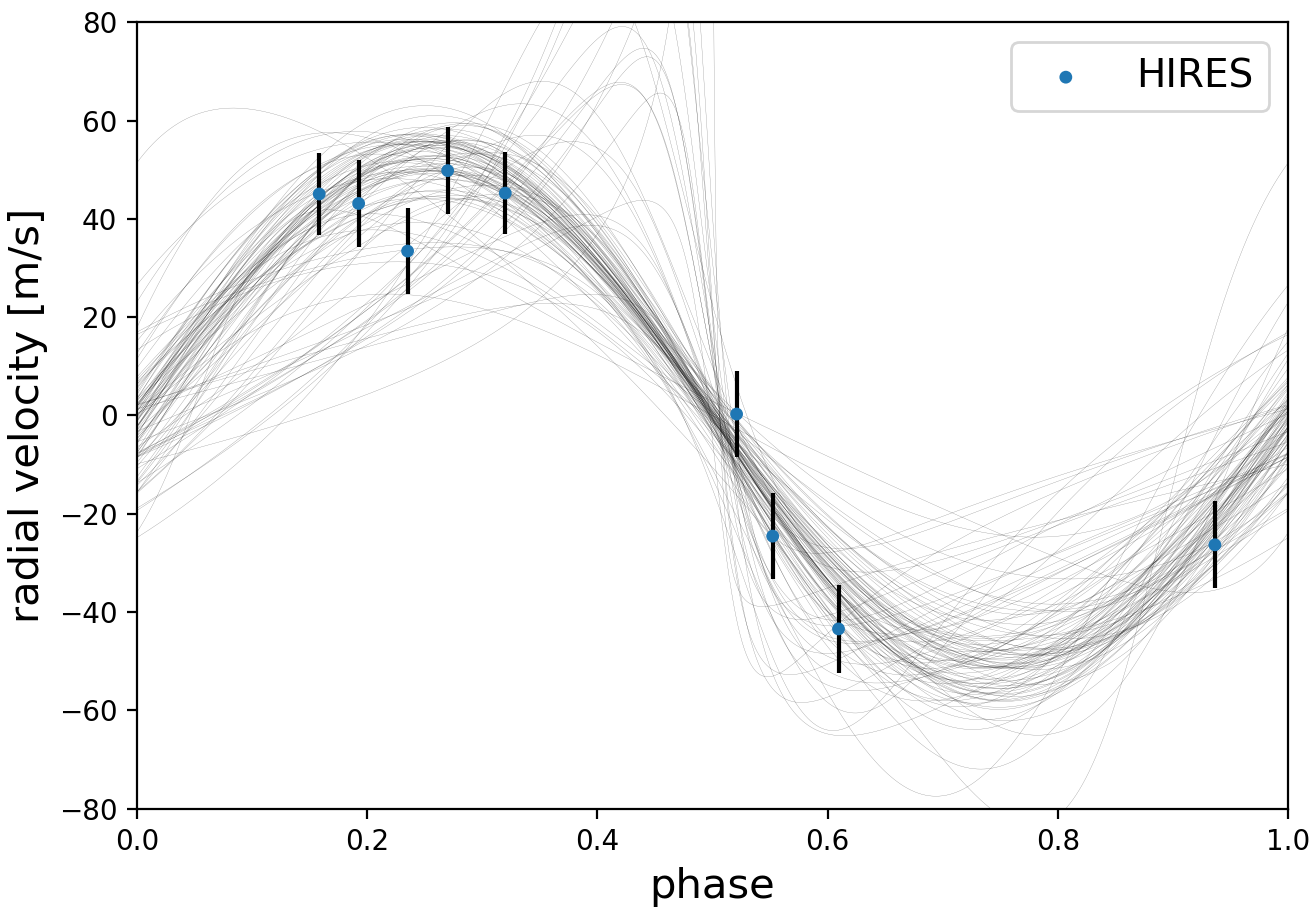}
    \caption{Same as Figure \ref{fig:lc_rv_2300} for \planettwo phase-folded at \periodtwo days. {\it Right:} Observations come from the Keck-I telescope on Maunakea.}
    \label{fig:lc_rv_2567}
\end{figure*}

\begin{figure*}[ht!]
    \centering
    \includegraphics[width=.49\textwidth]{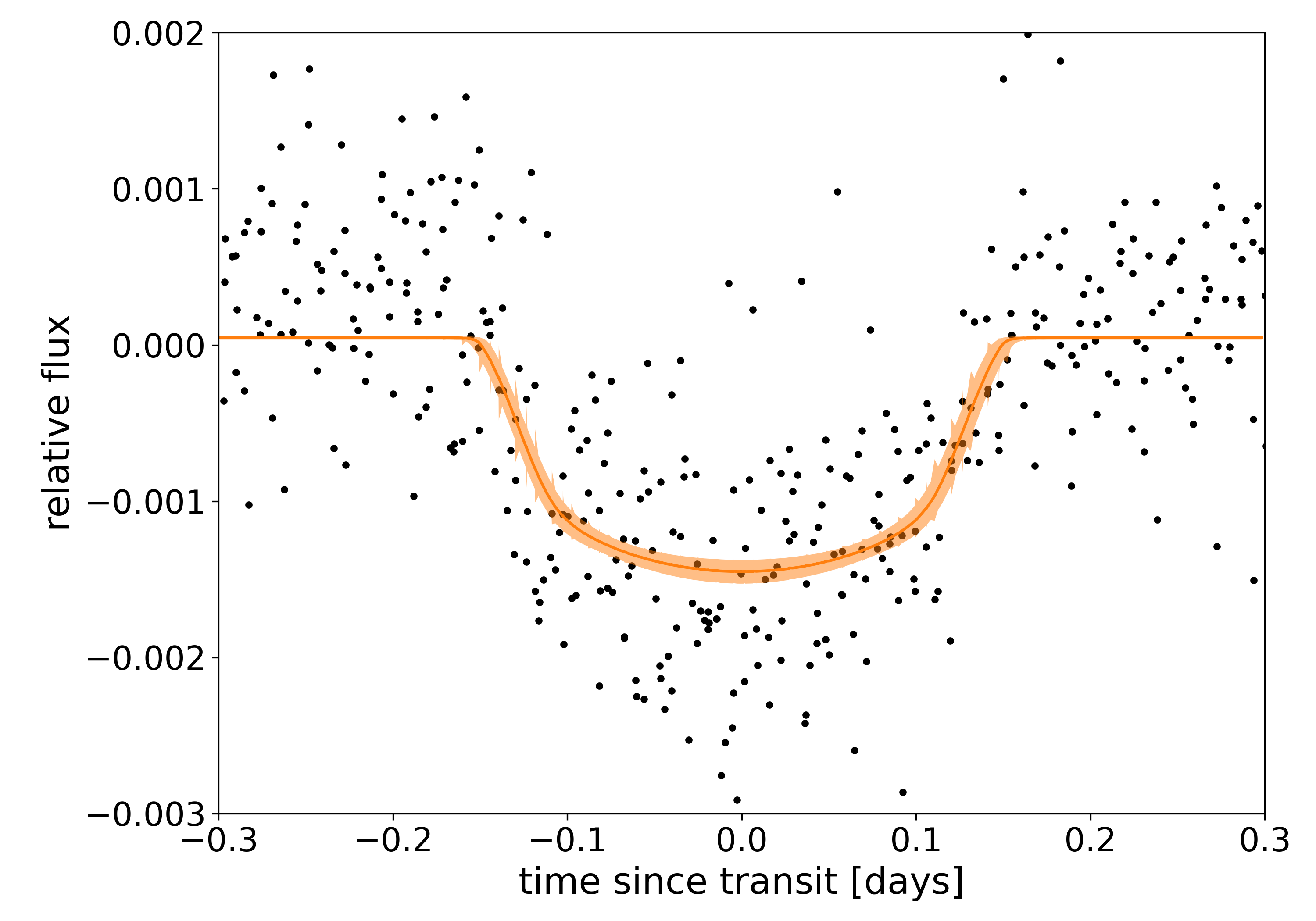}
    \includegraphics[width=.49\textwidth]{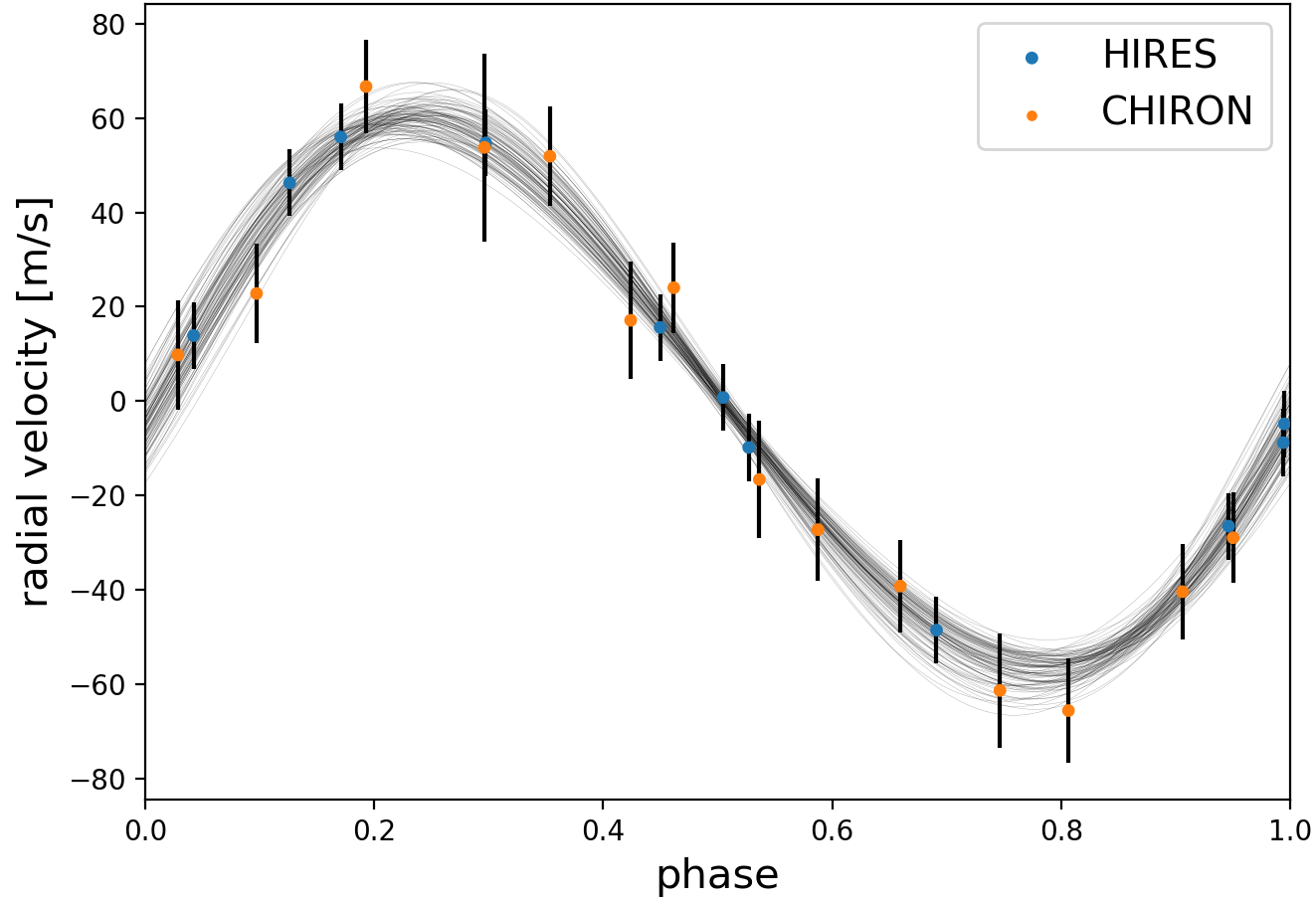}
    \caption{Same as Figure \ref{fig:lc_rv_2300} for \planetthree, phase-folded at \periodthree days. {\it Right:} Observations come from the Keck-I telescope on Maunakea (blue) and the SMARTS-1.5m telescope at CTIO (orange).}
    \label{fig:lc_rv_3488}
\end{figure*}

We present our best fit models to the light curve and radial velocity data for \hoststar in Figure \ref{fig:lc_rv_2300} and Table \ref{table:planet}. We investigate the additional out-of-transit light curve variability of \hoststar in Figure \ref{fig:beermodel}. We then present our best fit models to the light curve and radial velocity data of \hoststartwo and \hoststarthree in Figures \ref{fig:lc_rv_2567} and \ref{fig:lc_rv_3488} and Tables \ref{table:planettwo} and \ref{table:planetthree} respectively. We find that despite the similar masses we have measured for the host stars in these systems, the planet masses, radii and densities cover a wide range which is not clearly correlated with stellar properties. %Furthermore, we see additional deviations from standard planet transit models in at least the \planet system, which we discuss in more detail below.

For \planettwo, we performed a joint radial velocity and transit fit using the `KSP SAP', `SAP FLUX' and \texttt{giants} flux values. We find a planet radius of 1.37 $\pm$ 0.07 R$_\mathrm{J}$ using the \giants\ light curve, 1.60 $\pm$ 0.07 R$_\mathrm{J}$ using the `SAP FLUX' light curve, and 1.41 $\pm$ 0.07 R$_\mathrm{J}$ using the `KSPSAP FLUX' light curve. We find a maximum range in planet radii determined by these fits of 0.24 R$_\mathrm{J}$, significantly larger than the uncertainty in planet transit depth estimated from any fit to the available data. Thus, we assume that flux contamination must be the dominant source of uncertainty in planet radius determination for this system. As the difference between median planet radius reported between the two light curves is significantly larger than the errors on planet radius from either individual light curve, we add the statistical errors of each light curve fit in quadrature and add those to the one half the difference between median values to get a planet radius uncertainty of 0.19 R$_\mathrm{J}$ for \planettwo. Future observations with facilities with better resolution capable of using smaller photometric apertures, such as those used by the LCOGT, will confirm the true transit depth of this system, more precisely constraining the radius of \planettwo.

\subsection{BEER Modelling with \texttt{starry}}

\begin{figure}[ht!]
    \centering
    \includegraphics[width=.5\textwidth]{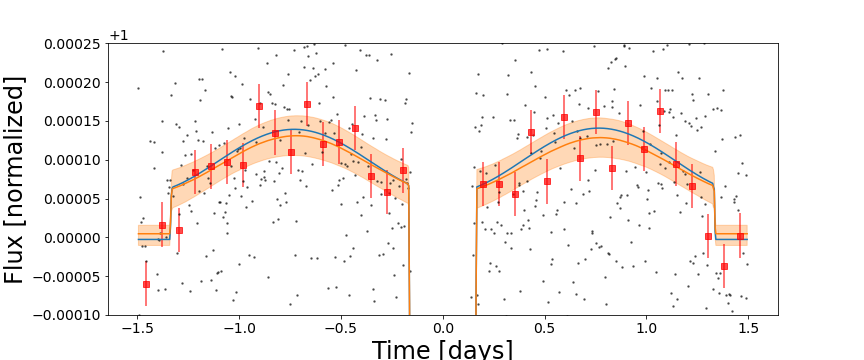}
    \caption{The flux light curve of \hoststar folded at a period of 2.99432 d, highlighting the out-of transit variability. The phase-folded, binned photometry is shown as black points. The initial \texttt{starry} model is shown in blue, and the best fit model is shown in orange, with 1-$\sigma$ confidence intervals highlighted. Ellipsoidal variation dominates the out-of-transit phase curve variability.}
    \label{fig:beermodel}
\end{figure}

\begin{table*}
\begin{center}
    \begin{tabular}{l c r}
        \hline 
        Parameter & Prior & Value \\
        \hline
        \rule{0pt}{3ex}\textit{Transit Fit Parameters} & & \\
        \rule{0pt}{3ex}Orbital period $P_{\text{orb}}$ [days] & $\log\mathcal{N}[2.9943, 0.0015]$ &  \period \\
        Transit epoch $t_0$ [BJD - 2457000] & $\mathcal{N}[1793.80652, 1.0]$ & \transittime \\
        Impact parameter $b$ & $P_\beta(e\in[0,1])^\text{(a)}$  & $0.52\pm0.08$\\
        Eccentricity $e$ & single-planet dist. from \citet{vaneylen2019} & $0.019 \pm 0.017$\\
        Argument of periastron $\Omega$ & $\mathcal{U}[-\pi, \pi]$& -0.341 $\pm$ 1.228 \\
        Limb-darkening coefficient $q_1$ & [0,2]$^\text{(b)}$ & $0.460 \pm 0.261$ \\
        Limb-darkening coefficient $q_2$ & [-1,1]$^\text{(b)}$ & $0.322 \pm 0.346$ \\
        \hline
        \rule{0pt}{3ex}\textit{Radial Velocity Fit Parameters} & & \\
        \rule{0pt}{3ex}Semi-amplitude $K$ [m/s] & $\mathcal{U}[0, 500]$ & $192.2 \pm 2.6$\\
        \hline
        \rule{0pt}{3ex}\textit{Derived Physical Parameters} & & \\
        \rule{0pt}{3ex}Planet radius $R_p$ $(R_\oplus)$ & $\mathcal{U}[0, 3]$ & \planetradius \\
        Planet mass $M_p$ $(M_J)$ & $\mathcal{U}[0, 300]$ & \planetmass \\
        \hline
   \end{tabular}
	 \caption{Fit and derived parameters for \planet. \textit{Note:} $^\text{(a)}$This parameterization is described by the Beta distribution in \citet{kipping2013b}. $^\text{(b)}$  Distributions follow correlated} two-parameter quadratic limb-darkening law from \citet{kipping2013}.
	 \label{table:planet}
\end{center}
\end{table*}

The folded light curve of \hoststar revealed not only a transit signal, but also corresponding out-of-transit variability at the same orbital period. To model the out of transit phase curve variations in this star, we adopt a beaming, ellipsoidal variability, and reflection (BEER) model \citep{faigler2011} to account for reflection of light off of the planet, ellipsoidal variations in the star, and Doppler beaming of the light assisted by the planet. These effects can be modeled as trigonometric functions:

\begin{equation}
M(t) = a_0 + \alpha \mathrm{cos}\Big(\frac{2\pi}{P_\mathrm{orb}} t\Big) + \beta \mathrm{cos}\Big(\frac{2\pi}{P_\mathrm{orb} / 2} t\Big) + \gamma \mathrm{sin}\Big(\frac{2 \pi}{P_\mathrm{orb}} t\Big) ,
\end{equation}
where $\alpha$ represents the reflection brightness, $\beta$ represents the amplitude of the ellipsoidal variability, and $\gamma$ represents the Doppler beaming strength. We find that we can produce a better fit to the data by including significant ellipsoidal variability in our model. Following the formulation of \citet{shporer2017}, we expect the semiamplitude of ellipsoidal variability $A_\mathrm{ellip}$ can be estimated using,

\begin{equation}
A_\mathrm{ellip} \propto \frac{M_p}{M_*} \Big(\frac{R_*}{a}\Big)^3.
\end{equation}

We model the phase curve variability of this system following the hot Jupiter phase curve example of \citet{luger2018} with the \texttt{starry} package. This model allows us to map the surfaces of both the host star and exoplanet as a sum of spherical harmonics, resulting in a flux model which is equivalent to the BEER model described above. This model also allows us to fit for the offset in the peak flux of the sinusoidal or spherical harmonic components of the flux, allowing for additional flexibility not accessible with a pure BEER model as described above. For our spherical harmonic model, we include a dipole and quadrupole mode at the orbital phase of the planet, which is equivalent to the reflection and ellipsoidal variation terms in the above equation. As the Doppler beaming term we expect from this system is $<$ 10 ppm (as determined by Eqn. 4 in \citet{shporer2017}), we do not fit for it in our model. We then fit for the coefficients of the dipole and quadrupole terms, as well as an overall variability strength and phase offset term, to model the light curve of \planet, folded at the planet orbital phase listed in Table \ref{table:planet} and binned the folded light curve with a bin size of 48 points. We illustrate our best fit model and 1-$\sigma$ confidence intervals in Figure \ref{fig:beermodel}.

 We measure the dipole coefficient to be consistent with 0, and a quadrupole coefficient corresponding to a semiamplitude $\approx$55 ppm in the \tess\ \texttt{giants} light curve. Assuming a model for ellipsoidal variation following the formulation of \citet{shporer2017}, we expect the ellipsoidal variation to be most strongly influenced by the scaled orbital separation of the system, as mentioned above. Extrapolating from the measured ellipsoidal variability of other \tess\ systems \citep{wong2020,wong2021}, we predict an ellipsoidal semiamplitude of 75 $\pm$ 15 ppm for this system, in good agreement with our detection.

%which is in very good agreement with what is observed in the \tess\ \texttt{giants} and QLP light curves.

%However, even for our best fit BEER model, we cannot remove all of the variability in our phase-folded light curve, finding qualitatively similar dips as were seen in out-of-transit phase variability of Kepler-91 \citep{lillo-box2014}. 

A more complicated BEER model which allows for eccentricity may be able to remove some additional noise, but due to the strong constraint on eccentricity from radial velocity followup ($e < 0.05$ at a 95\% confidence interval) we do not allow eccentricity in our out-of-transit phase variability model. The strongest out-of-transit phase signal is clearly due to a quadrupole-like variation such as ellipsoidal variability. In addition, we measure a significant offset in phase for the out of transit sinusoidal variation. Allowing the ellipsoidal variability to lag the transit in phase produces a better fit to the data. % In addition, we find an unexplained brightening in flux at a phase of $\sim$0.7. This could potentially be described by a more eccentric planet model, which again would be inconsistent with the radial velocity data. 

%The ellipsoidal variability we observe in the \planet system indicates a slight eccentricity. This is unexpected for such a short period planet \citep{villaver2009,villaver2014}. However, the radial velocity measurements indicate that the eccentricity of this system $e <$ 0.05 at a 95\% confidence interval. 

No significant out-of-transit variability can be seen for \planettwo or \planetthree. Recent 10-minute cadence data available for \planetthree gives evidence for a relatively short ($\approx$ 6 hr) transit duration, indicative of either a high impact parameter transit or eccentricity in the planet's orbit, in agreement with the planet parameters presented here. Combined with previous observations and radial velocity data of \planetthree, a low to moderate eccentricity is more likely for this planet ($e < 0.2$) and thus this suggests transit may have a high impact parameter.

\begin{table*}
\begin{center}
    \begin{tabular}{l c r}
        \hline 
        Parameter & Prior & Value \\
        \hline
        \rule{0pt}{3ex}\textit{Transit Fit Parameters} & & \\
        \rule{0pt}{3ex}Orbital period $P_{\text{orb}}$ [days] & $\log\mathcal{N}[2.923, 0.001]$ &  \periodtwo \\
        Transit epoch $t_0$ [BJD - 245700] & $\mathcal{N}[1765.067, 0.05]$ & \transittimetwo \\
        Impact parameter $b$ & $P_\beta(e\in[0,1])^\text{(a)}$ & $0.24\pm0.18$\\
        Eccentricity $e$ & single planet dist. from \citet{vaneylen2019} & $0.068 \pm 0.079$\\
        Argument of periastron $\Omega$ & $\mathcal{U}[-\pi, \pi]$& 0.285 $\pm$ 2.120 \\
        Limb-darkening coefficient $q_1$ & [0,2]$^\text{(b)}$ & $0.258 \pm 0.165$ \\
        Limb-darkening coefficient $q_2$ & [-1,1]$^\text{(b)}$ & $0.535 \pm 0.263$ \\
        \hline
        \rule{0pt}{3ex}\textit{Radial Velocity Fit Parameters} & & \\
        \rule{0pt}{3ex}Semi-amplitude $K$ [m/s] & $\mathcal{U}[0, 300]$ & $55.6 \pm 4.1$\\
        \hline
        \rule{0pt}{3ex}\textit{Derived Physical Parameters} & & \\
        \rule{0pt}{3ex}Planet radius $R_p$ $(R_\oplus)$ & $\mathcal{U}[0, 3]$ & \planetradiustwo \\
        Planet mass $M_p$ $(M_J)$ & $\mathcal{U}[0, 300]$ & \planetmasstwo \\
        \hline
   \end{tabular}
	 \caption{Fit and derived parameters for \planettwo. \textit{Note:} $^\text{(a)}$This parameterization is described by the Beta distribution in \cite{kipping2013b}. $^\text{(b)}$ Distributions follow correlated} two-parameter quadratic limb-darkening law from \cite{kipping2013}.
	 \label{table:planettwo}
\end{center}
\end{table*}

\begin{table*}
\begin{center}
    \begin{tabular}{l c r}
        \hline 
        Parameter & Prior & Value \\
        \hline
        \rule{0pt}{3ex}\textit{Transit Fit Parameters} & & \\
        \rule{0pt}{3ex}Orbital period $P_{\text{orb}}$ [days] & $\log\mathcal{N}[6.203, 0.01]$ &  \periodthree \\
        Transit epoch $t_0$ [BJD - 2457000] & $\mathcal{N}[1521.56, 0.1]$ & \transittimethree \\
        Impact parameter $b$ & $P_\beta(e\in[0,1])^\text{(a)}$  & $0.858\pm0.025$\\
        Eccentricity $e$ & single-planet dist. from \citet{vaneylen2019} & $0.09 \pm 0.05$\\
        Argument of periastron $\Omega$ & $\mathcal{U}[-\pi, \pi]$& -1.220 $\pm$ 0.585 \\
        Limb-darkening coefficient $q_1$ & [0,2]$^\text{(b)}$ & $0.68 \pm 0.48$ \\
        Limb-darkening coefficient $q_2$ & [-1,1]$^\text{(b)}$ & $-0.01 \pm 0.41$ \\
        \hline
        \rule{0pt}{3ex}\textit{Radial Velocity Fit Parameters} & & \\
        \rule{0pt}{3ex}Semi-amplitude $K$ [m/s] & $\mathcal{U}[0, 300]$ & $59.6 \pm 3.0$\\
        \hline
        \rule{0pt}{3ex}\textit{Derived Physical Parameters} & & \\
        \rule{0pt}{3ex}Planet radius $R_p$ $(R_\oplus)$ & $\mathcal{U}[0, 3]$ & \planetradiusthree \\
        Planet mass $M_p$ $(M_J)$ & $\mathcal{U}[0, 100]$ & \planetmassthree \\
        \hline
   \end{tabular}
	 \caption{Fit and derived parameters for \planetthree. \textit{Note:} $^\mathrm{(a)}$This parameterization is described by the Beta distribution in \citet{kipping2013b}. $^\mathrm{(b)}$Distributions follow correlated} two-parameter quadratic limb-darkening law from \citet{kipping2013}.
	 \label{table:planetthree}
\end{center}
\end{table*}

%\section{Discussion} \label{sec:discussion}

%\subsection{Evolved Planet Population Statistics}

%The late-stage evolution of close-in planetary systems is poorly understood, due to the relatively small numbers of systems known. Here we discuss what effect planet discoveries have for understanding the population of planets hosted by evolved stars, focusing specifically on planet (re-)inflation, star-planet interaction, and stellar structure, and then explore the value of additional ground-based and space-based followup, and potential for statistical constraints of this planet population.

\section{Planet Radius Inflation} \label{sec:inflation}

\begin{figure}[ht!]
    \centering
    \includegraphics[width=.5\textwidth]{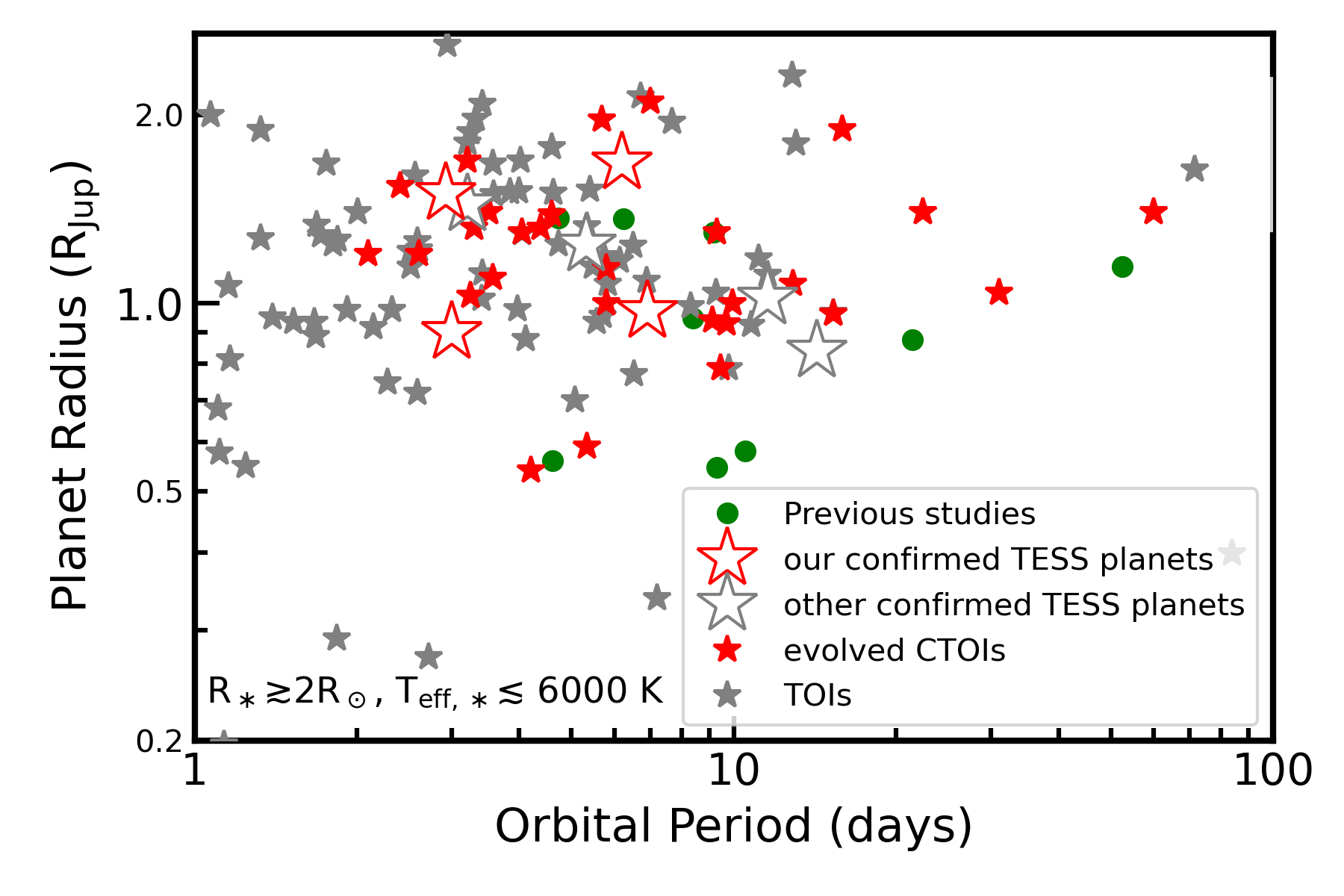}
    \caption{Orbital period versus planet radius for confirmed planets and new candidates transiting evolved (R$_*$ $>$ 2 R$_\odot$, T$_\mathrm{eff}$ $<$ 6000 K) stars. Planets known around evolved stars before the launch of \tess are shown in green. Those confirmed by \tess\ are shown as the largest symbols. Additional community-flagged planet candidates found by \tess\ are shown as small red stars, and \tess\ Objects  of Interest (TOIs) are shown in gray.}
    \label{fig:evolved_pop}
\end{figure}

\begin{figure*}[ht!]
    \centering
    \includegraphics[width=\textwidth]{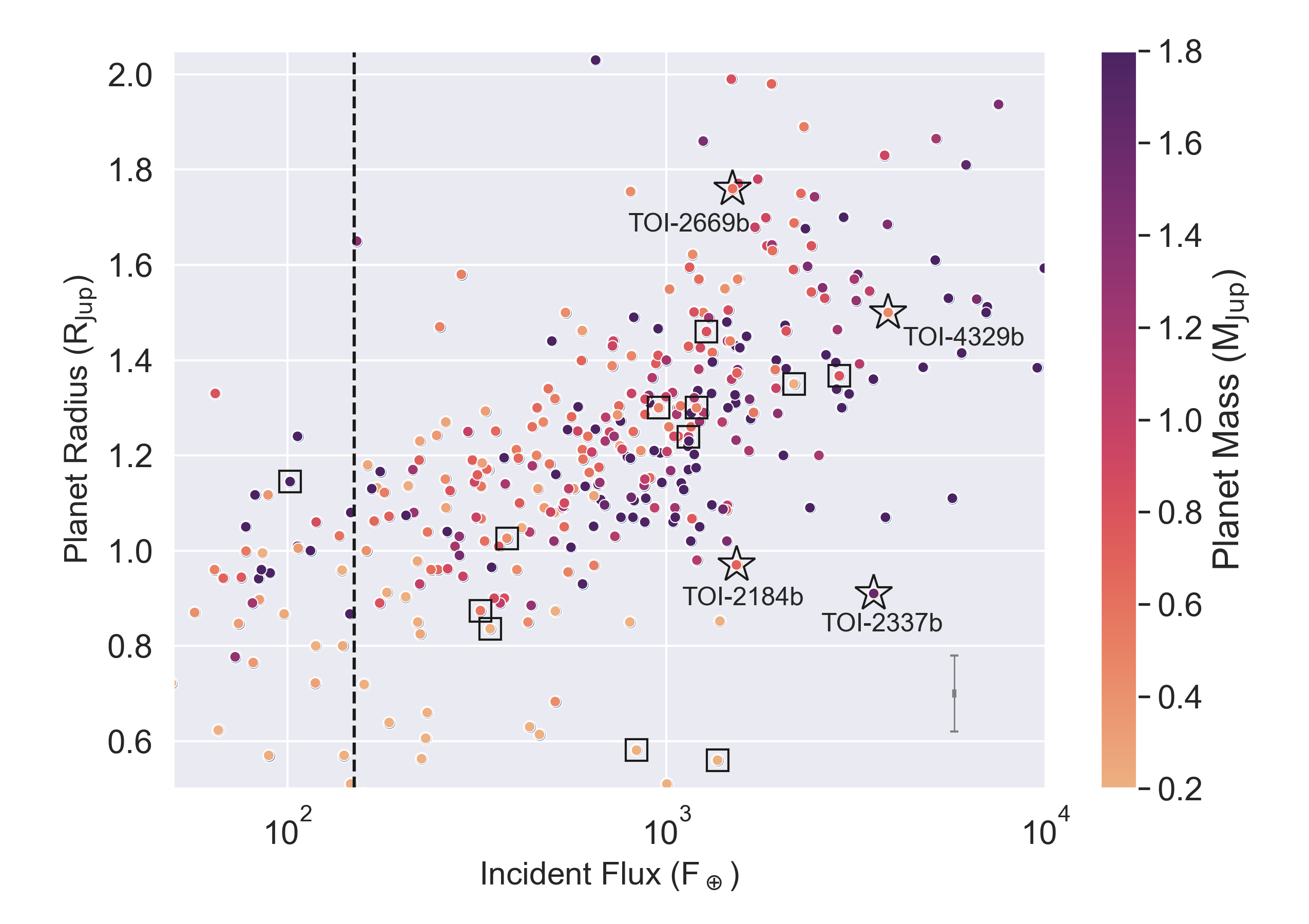}
    \caption{Planet radius vs. incident flux for all known hot Jupiters. The black line corresponds to the observed threshold for planet inflation \citep[150 F$_\oplus$;][]{demory2011}. Color corresponds to planet mass. Typical errors for the data are shown by the gray point in the lower right hand corner of the plot. Previously known evolved (R$_*$ $>$ 2 R$_\odot$, T$_\mathrm{eff}$ $<$ 6000 K) systems have been highlighted with squares, whereas the new systems confirmed here and TOI-2184b, found in the same planet search, have been highlighted as stars and labeled on this plot. \planettwo and \planetthree are in good agreement with the larger population trend of known hot Jupiters, while \planet is an outlier. Evolved systems do not appear to follow a straightforward planet mass-radius-flux relationship, suggesting other factors may be important in the late-stage inflation of planets.}
    \label{fig:rad_v_flux}
\end{figure*}

Previous to our \tess\ survey, only seven planets had been confirmed to be transiting evolved {/bf (T$_{\mathrm{eff}} <$ 6000 K, R $>$ 2 R$_\odot$)} stars. These systems showed promise for solving mysteries of late-stage planet inflation and orbital evolution, but small numbers and undersampled parameter space has made determining population-wide characteristics difficult. With our additional discoveries (including TOI-2184b from \citet{saunders2021}), our \tess\ survey has now increased the number of confirmed planets in this population by almost 50\%, and has revealed a new regime of short period hot Jupiters that have not yet inspiraled into their evolved host stars. In addition, our survey has revealed a number of other similar planet candidates which will be suitable for confirmation or at least ground-based followup in the near future. Measurements of the masses and eccentricities of these and similar systems will provide new constraints on planetary inflation and evolution, and the rate of inflation in these candidates is already providing new constraints on the mechanisms and timescales for re-inflation that were not possible earlier. We illustrate previously known and new planets, as well as new planet candidates orbiting evolved stars, in Figure \ref{fig:evolved_pop}.

We illustrate planet radius as a function of incident flux in Figure \ref{fig:rad_v_flux}, highlighting the hot Jupiter population. Color indicates planet mass, and we illustrate the planets confirmed by this work as stars, and other planets around evolved stars as squares. A strong correlation between incident flux and planet radius can be seen for planets $\gtrsim$ 0.5 M$_\mathrm{Jup}$ \citep{demory2011,lopez2016,thorngren2018, sestovic2018}. However, a number of outliers to these trends can also be seen--for example, \planet is the smallest planet known with a mass $>$ 0.3 M$_\mathrm{Jup}$ and incident flux $>$ 2000 F$_\oplus$, TOI-2184b is similarly significantly underinflated, and \planettwo appears somewhat underinflated for its mass and incident flux while \planetthree appears relatively overinflated. This indicates that planet re-inflation at late times may be driven by a combination of different factors, but implies that planets can become more inflated at late evolutionary stages \citep{lopez2016, grunblatt2017, grunblatt2019}. Atmospheric stripping could play a role at such small orbital separations, changing the potential radius and composition of the planet significantly \citep{bell2019, baxter2021,swain2021}.

Relative to other known evolved systems, the incident flux on \planet is quite high. Thus assuming a direct correlation between planet radius and incident flux, it would be expected that this planet is inflated, yet it is not. As seen in Fig. \ref{fig:rad_v_flux}, the incident flux received by \planet is greater than that typically received by similarly-sized hot Jupiters---clustered to the left of \planet---by roughly a factor of 2. On the contrary, both \planettwo and \planetthree appear to have inflated radii, and smaller masses than \planet. 

When taking all evolved systems into account, it appears that they are not evenly distributed among the larger population of inflated Jupiters, but instead seem to prefer relatively low rates of inflation regardless of planet mass. This suggests that for evolved stars, a straightforward mass-radius-flux power-law relation may not be sufficient to describe the observed population, as flux integrated over time and change in flux is also relevant for these systems \citep{weiss2013, sarkis2021}. These flux changes are highly dependent on stellar evolution as well as planetary orbital dynamics. Alternatively, \planet may simply be an outlier from the typical population of hot Jupiters that are described well by a mass-radius-flux relation, due to mass loss or an unusually massive core.

% Evolved stars seem to host both some of the least inflated as well as some of the most inflated planets, all of which have relatively high incident fluxes and eccentricities likely below 0.1. 

These systems can be interpreted as evidence for rapid re-inflation \citep{thorngren2021}, which could result in the large difference in planet radii among these particularly high equilibrium temperature planets. The vastly different planet radii could be due to different strengths and depths of heat dissipation processes for the planets studied here \citep{komacek2020}, but uncertainties in the bulk metallicity and the migration history of these planets may also play a role. The orbits currently observed for these planets may have been reached through circularization induced by stellar evolution \citep{villaver2009,villaver2014}, which could also result in time-dependent internal heating and radius inflation due to the changing planetary orbit.

%If evidence can be found for different inflation mechanisms in different systems, this may have implications for planet inflation as a function of stellar evolutionary state.

\section{Eccentricity Analysis}

% \begin{figure*}[ht!]
%     \centering
%     \includegraphics[width=.98\textwidth]{eccentricityfig.png}
%     \caption{Orbital period vs. planet eccentricity for the dwarf and giant star populations. Dwarf planets are marked by the dark black circles, and giant planets are the red circles. The red stars represents the new planet candidates presented here. The systems studied here seem to follow trend identified in \citet{grunblatt2018}, where planets orbiting giant stars at short periods prefer moderately eccentric orbits.}
%     \label{fig:per_rad}
% \end{figure*}

\citet{grunblatt2018} showed that giant planets orbiting giant stars at periods $<$30 days have average eccentricities $e>0.1$. However, at the shortest orbital periods ($<$5 days), even planets around evolved stars appear to have largely circular orbits. The strong upper limits on eccentricities of the orbits in the evolved systems presented here suggests that these systems have largely completed orbital circularization and inspiral as described in \citet{villaver2014}. Constraints on orbital eccentricities will constrain both planet engulfment and stellar structure models \citep{weinberg2017, sun2018, soaresfurtado2020}. We find no evidence for significant eccentricity in observations of any three of our systems. Both \planet and \planettwo have eccentricities $e < 0.05$, while the longer-period \planetthree has an eccentricity potentially inconsistent with zero, yet significantly smaller than $0.15$. Given the remarkably short orbital periods of these systems relative to other transiting evolved systems, these results are in strong agreement with the findings of \citet{grunblatt2018}, but a wider range of eccentricities and periods are needed to strongly support the existence of a correlation between orbital eccentricity and period for transiting planets in evolved giant planet systems.

%This implies additional measurements are necessary to further constrain star and planet evolution in these systems. 

% \begin{figure}[ht!]
%     \centering
%     \includegraphics[width=.45\textwidth]{incflux_093020.png}
%     \caption{Incident flux received by \planet as a function of evolutionary state of the star. The current evolutionary state of \hoststar is marked by the green region. \planet has received a flux significantly above the planet inflation threshold for its entire main sequence and post main sequence lifetime.}
%     \label{fig:rad_v_flux}
% \end{figure}

% \begin{figure}[ht!]
%     \centering
%     \includegraphics[width=.45\textwidth]{mass_radius.pdf}
%     \caption{Planet mass versus planet radius for confirmed exoplanets. Confirmed planets discovered by \tess are shown as black points while planets discovered by other telescopes are shown in gray.}
%     \label{fig:mass_radius}
% \end{figure}

% \begin{figure}[ht!]
%     \centering
%     \includegraphics[width=.45\textwidth]{mass_radius_tracks.pdf}
%     \caption{Planet mass versus planet radius for confirmed exoplanets, zoomed in on the Hot Jupiter parameter space. Confirmed planets discovered by \tess are shown as black points while planets discovered by other telescopes are shown in gray. The colored lines show various models for planetary density by \cite{freedman2014}. These models do not account for planetary inflation, }
%     \label{fig:mass_radius}
% \end{figure}

\section{Potential JWST Followup Observations}

\begin{figure*}[ht!]
    \centering
    \includegraphics[width=.85\textwidth]{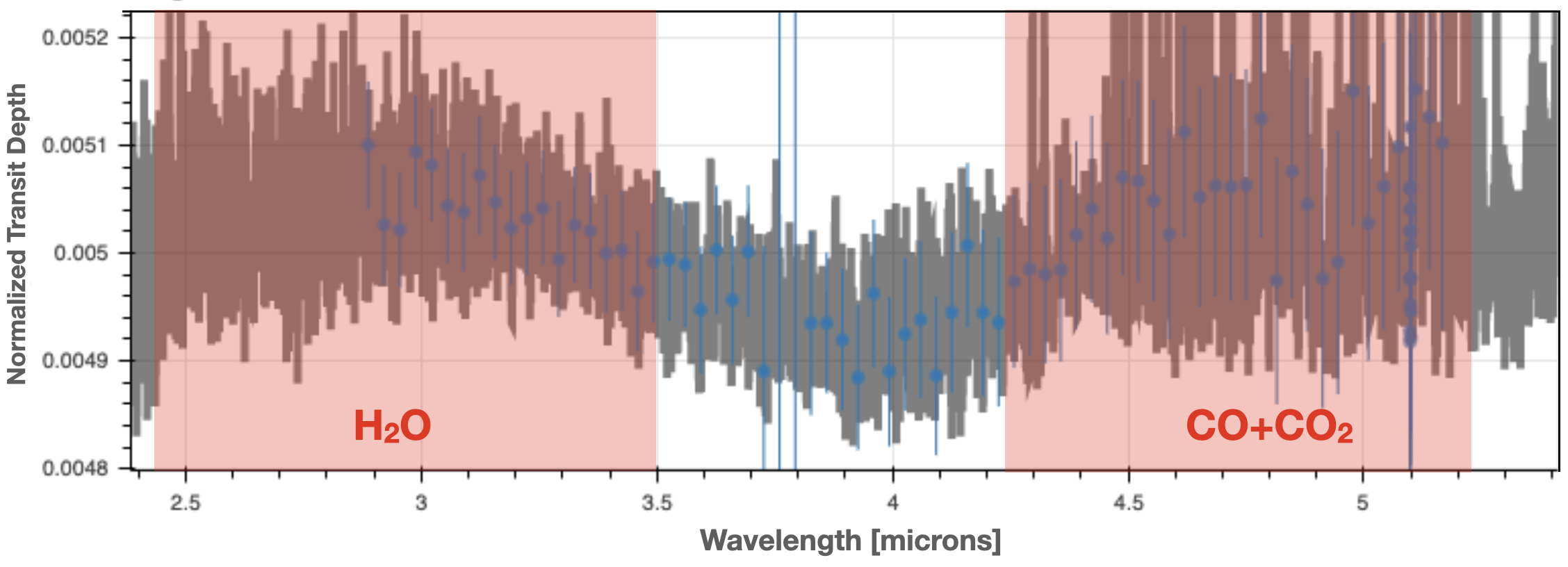}
    \caption{Transit depths for \planettwo as measured by JWST's NIRSPEC instrument (blue points) compared to a default planet model atmosphere with equilibrium chemistry and medium cloud cover (gray), at the precision expected for one transit observation \citep{batalha2017}. The regions of the spectrum sensitive to water and carbon species in the planet's atmosphere have been highlighted. The strength of these features have implications for the atmospheric composition and mixing in this planet, and the inflation of post-main sequence planets in general.}
    \label{fig:simspec}
\end{figure*}

Little is understood about the evolution of planetary atmospheres over time, and no observations of planetary atmospheres have been conducted for planets orbiting evolved stars. Due to their particularly short orbits, the systems introduced here are more well-suited for atmospheric characterization than other planets orbiting evolved stars.

Observations with the James Webb Space Telescope (JWST) will enable the comparison of these evolved planets directly to hot and ultra-hot Jupiters orbiting main sequence and pre-main sequence stars. As it is believed that the hot Jupiters we find on short orbits around evolved stars may have begun their lives on much larger orbits around young stars, comparison between young, main sequence, and evolved planet transmission spectra could reveal evolution of planetary atmospheres.

We find that \planettwo has a transmission spectroscopy metric \citep[TSM;][]{kempton2018} value of $\sim$80, making it the best candidate hosted by a $>$2 R$_\odot$ star for JWST followup. The shallower transit depths of \planet and \planetthree make them less suitable to such followup, though \planetthree may be amenable to transmission spectroscopy with precision roughly 70\% as good as that of \planettwo (TSM $\sim$ 55).%, as well as in the third tier of \tess\ transmission spectroscopy targets larger than Neptune

Figure \ref{fig:simspec} shows the relative transit depth expected to be receovered at different wavelengths by JWST NIRSPEC for \planettwo, assuming the default planet atmospheric model of \texttt{PandExo}, with equilibrium chemistry and medium cloud cover and the measured planet mass and temperature \citep{batalha2017}. The wavelength range probes water and carbon dioxide features, both which are clearly visible given the relative brightness and precise mass of the planet.

These observations are predicted to constrain the carbon-to-oxygen ratio in the atmosphere of \planettwo, informing formation and migration scenarios for these planets as has been done previously for planets orbiting main sequence stars \citep{line2014, benneke2015}. The relationship between planet mass and atmospheric metallicity of \planettwo will be particularly informative in understanding the formation and migration of this system \citep{welbanks2019}. The differences between the chemical abundances of this planet and of planets in main sequence systems may reveal how late stage stellar evolution impacts the evolution of planetary atmospheres.

Recent observations have suggested a distinction in the atmospheric profiles of hot and ultra-hot Jupiters \citep{baxter2021}. Additionally, transmission spectra probe atmospheric composition, which can inform what fraction of the atmosphere is primordial versus accreted at later stages of planet development. Atomic and molecular abundances measured from transmission spectroscopy may also probe where the planet originally formed and at what point in its lifetime it moved to its current orbit \citep{oberg2011,dawson2018}.

In addition, observations of the strengths of atomic and molecular features will constrain the strength of vertical mixing and resulting amount of chemical disequilibrium in this planet's atmosphere, and test theories of planet inflation and atmospheric mixing, which are predicted to change for planets below and above 1000 K \citep{komacek2019, baxter2021}. As \planettwo likely crossed this threshold in its near past, the level of chemical disequilibrium seen here will be informative regarding the timescales of these mixing processes. %Only JWST has the necessary precision to probe these chemical signatures of this planet's atmosphere.

\begin{figure*}[ht!]
    \centering
    \includegraphics[width=.95\textwidth]{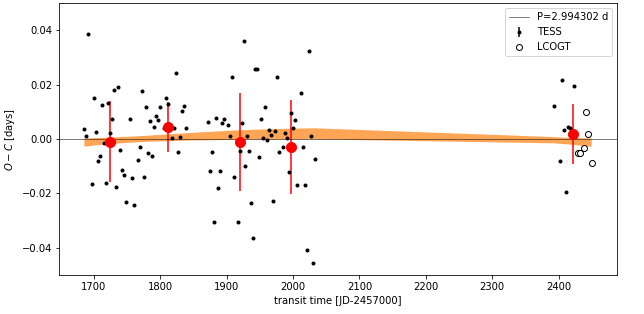}
    \caption{Transit times of \planet from \tess and LCOGT Teide observations as measured by the \texttt{exoplanet} TTVOrbit package. We use a quadratic formula to describe the evolution of transit times, and do not detect any clear orbital decay in this system. Although there is still a great deal of scatter in the measured transit times from our current data for this system, additional \tess\ extended Mission data has only just become available, and should constrain the orbital decay rate of this system to significantly higher precision by the end of next year.}
    \label{fig:ominusc}
\end{figure*}

\section{Orbital Decay of the Evolved Planet Population}\label{sec:orbdec}

%There have been very few close-in planets discovered around massive evolved red giant stars, making \planet an important discovery to help understand the theory for the formation and evolution of post-main sequence planetary systems. The evolutionary stage of \hoststar also makes this a compelling candidate for measuring tidal interactions between planets and their evolving host stars. 

The expected tidal interaction between hot gas giant planets and evolved host stars is expected to result in rapid orbital decay and eventual engulfment of the planet. However, orbital decay has only been measured in one system to date, which was significantly less evolved than the systems studied here. By constraining the rate of orbital decay in these systems, we can measure the strength of star-planet tidal interactions and their dependence on star and planet properties.

Based on the \tess\ and ground-based data currently available for this system, we measure the individual transit times as observed by \tess\ and LCOGT and fit them using a quadratic model through the \texttt{exoplanet} TTVOrbit package. We choose to include only the LCOGT observations from Teide observatory, as they give more consistent and numerous transit times than can be derived from the McDonald Observatory observations, which were more strongly affected by poor weather conditions. We illustrate our measured transit times and best-fit quadratic model in Figure \ref{fig:ominusc}.

Our best-fit model does not significantly prefer orbital decay to describe this system. Most importantly, the transit times measured by LCOGT and the \tess\ Extended Mission appear to be in strong agreement with the expected transit times for this system, implying that any orbital decay in this system has not changed the ephemerides of transit by $>$0.02 d, and thus is not yet measurable. This implies that the change in period of this system is $<$1 second per year, and following the ``constant phase lag" formulation of \citet{goldreich1966} used in recent orbital decay detection, this corresponds to $Q'_\star > 2 \times 10^4$, in agreement with what has been found for other systems where orbital decay has been constrained \citep{chontos2019,yee2020,patra2020}. A longer baseline of data with 2-minute cadence \tess\ data for \planet will provide tighter constraints on the orbital decay rate of this system. Given the more limited data sets available for \planettwo and \planetthree, we do not attempt to measure orbital decay in these systems.

%prefers orbital decay for this system, although at less than a 68\% confidence level. We extrapolate our best-fit orbit decay model to the times at which additional observations will be made by \tess, and find that if our model is correct, transits should be arriving $\approx$35 minutes early by the start of \tess\ Extended Mission observations of this target, which should be easily measurable with the 2-minute (and 10-minute) cadence data that will be made available for \planet. We do not see any evidence for orbital decay in the other systems introduced here.

%The lack of eccentricity in the shortest period planets is to be expected from the theory of planetary inspiral \citep{villaver2014, sun2018, macleod2018, soaresfurtado2020}. 

\begin{figure*}[ht!]
    \centering
    \includegraphics[width=\textwidth]{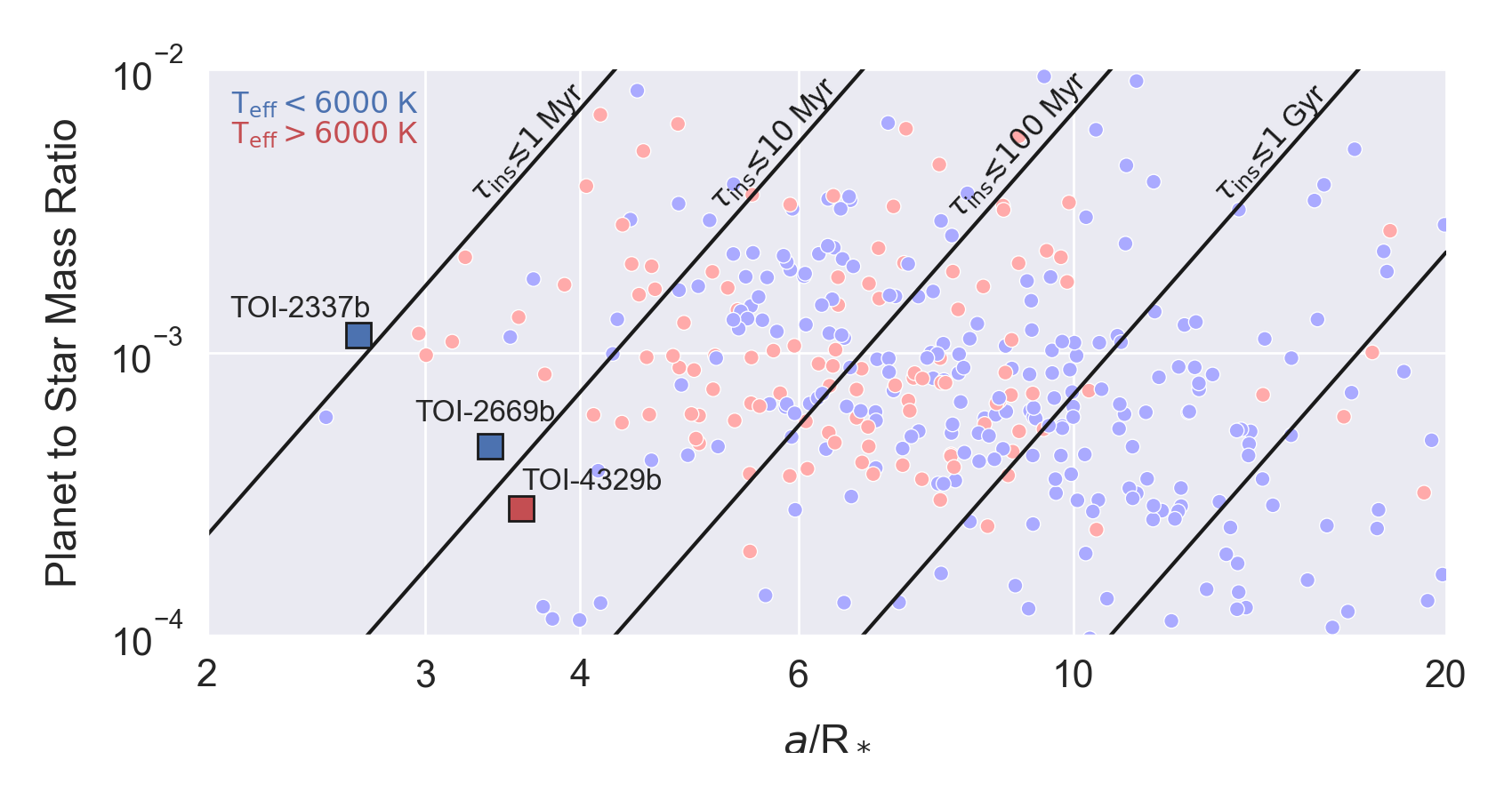}
    \caption{Semimajor axis divided by stellar radius, versus planet to star mass ratio for confirmed planets. Orbital decay timescales decrease toward the upper left of this plot, where black diagonals correspond to theorized rates of orbital decay, where the leftmost line corresponds to a decay timescale of 10$^6$ years, and each following line increases by a factor of 10. Blue points have stellar effective temperatures $<$6000 K as reported by the NASA Exoplanet Archive, while red points represent planets around hotter stars. The planets confirmed by this work are shown as squares with black outlines, and are populating relatively sparse regions of parameter space on this plot that correspond to rapid orbital decay. In particular, \planet may be experiencing the fastest rate of orbital decay of any planet known to date.}
    \label{fig:orbdec_pop}
\end{figure*}

Figure \ref{fig:orbdec_pop} illustrates the population of known planets, highlighting those planets which are most likely to be experiencing strong orbital decay, as well as decay rates predicted using the equilibrium tide model of \citet{goldreich1966}. The planets with the smallest relative orbital separations and highest masses relative to their stars decay most quickly, and can be found in the upper left hand corner of this plot. We have illustrated the new planets found by this survey as squares on this plot. These planets are among some of the best candidates for detecting orbital decay. In particular, \planet is predicted to be the most rapidly decaying planet known to date. In addition, \planet also orbits a relatively cool star, which is expected to increase the speed of its orbital decay due to more rapid tidal dissipation in \hoststar's thick outer convective envelope \citep{patra2020}. Additional \tess\ data from the Extended \tess\ Mission will be essential to constraining the transit ephemerides and thus the orbital properties and stellar tidal quality factor for this system.

% \begin{figure}[ht!]
%     \centering
%     \includegraphics[width=.5\textwidth]{ms18_update.png}
%     \caption{Semimajor axis vs. log($g$) for confirmed planets, taken from \citet{sun2018}. Planets confirmed around evolved stars by \emph{Kepler} are shown as black stars. Planets confirmed by this work are shown as red stars. Kepler-91b and \planet are both beyond the critical semimajor axis value at which dynamical tides overpower equilibrium tides and runaway inspiral begins, implying that both planets will be destroyed.}
%     \label{fig:sunupdate}
% \end{figure}

\begin{figure}[ht!]
    \centering
    \includegraphics[width=0.5\textwidth]{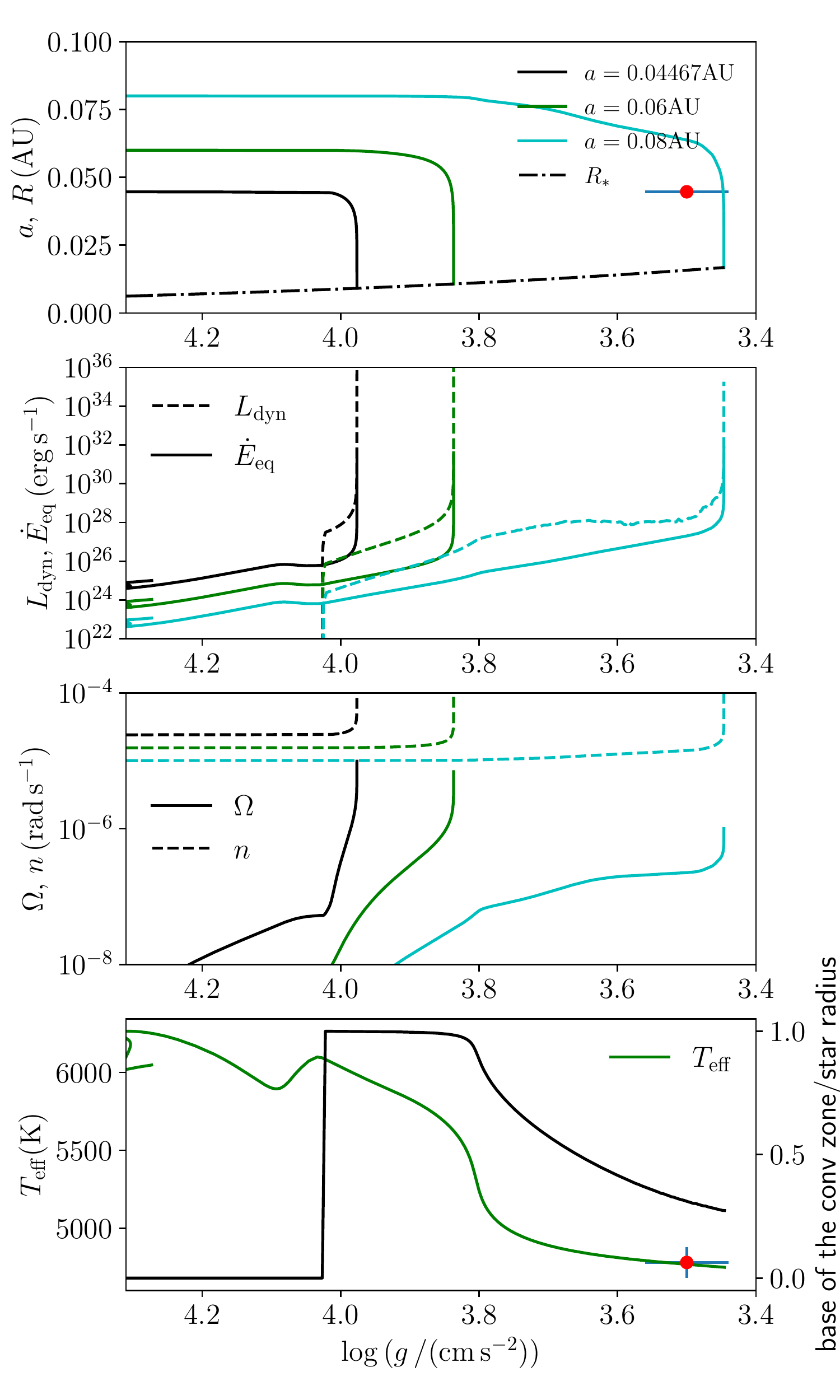}
    \caption{Top: Semimajor axis vs. log($g$), as well as evolutionary models for \planet. Second panel: Energy generated by dynamical and equilibrium tides during the inspiral of \planet, as a function of log($g$). At all evolutionary states, dynamical tides appear to dominate over equilibrium effects. Third panel: Rotational frequency vs. log($g$), where $\Omega$ indicates stellar rotation and $n$ indicates orbital period. The orbital period is significantly shorter than the stellar rotation at all relevant log($g$) values. Bottom panel: Stellar structure and effective temperature as a function of log($g$). As the host star \hoststar\ evolves, the convection zone first disappears from the core of the star, then deepens from the surface while the effective temperature decreases.}
    \label{fig:planettides}
\end{figure}

% \begin{figure*}[ht!]
%     \centering
%     \includegraphics[width=.85\textwidth]{inner_comp.pdf}
%     \caption{Upper Left: Semimajor axis vs. log($g$), as well as evolutionary models for \planet. Lower left: Energy generated by dynamical and equilibrium tides during the inspiral of \planet, as a function of log($g$). At all evolutionary states, dynamical tides appear to dominate over equilibrium effects. Upper right: Rotational frequency vs. log($g$), where $\Omega$ indicates stellar rotation and $n$ indicates orbital period. The orbital period is significantly shorter than the stellar rotation at all relevant log($g$) values. Lower right: Stellar structure and effective temperature as a function of log($g$). As the host star \hoststar\ evolves, the convection zone deepens and the effective temperature decreases.}
%     \label{fig:stellinter}
% \end{figure*}

Based on the orbital properties of these systems, and theoretical assumptions about binary system evolution \citep{1977A&A....57..383Z,hut1981,1989A&A...220..112Z,1989A&A...223..112Z}, we can infer the initial orbital properties, relevant energy dissipation rates, and stellar structure constraints that could not be probed in other systems. Figure \ref{fig:planettides} illustrates the strength of various tides and the expected evolutionary tracks for \planet and \hoststar. The tidal quality factor of \hoststar, $Q'_\star$ is highly dependent on the evolutionary processes. Assuming a tidal quality factor for this star of 2 $\times$ 10$^5$ as measured in WASP-12b \citep{yee2020,turner2021}, we can predict the orbital decay of \planet\ by assuming an initial semimajor axis and predicting the orbital evolution of the system. As the combination of orbital period and stellar surface gravity is the most extreme for \planet,  here we approach the dynamical analysis of the \hoststar system via the numerical calculation of the tidal energy dissipation rate over time instead of assuming a constant value for $Q'_\star$.

To estimate the order of magnitude of the rate of change in the semimajor axis and orbit period, we follow the method used in \citealt{weinberg2017,sun2018} for close circular exoplanet systems. The stellar model is constructed with the Modules for Experiments in Stellar Astrophysics (MESA; Version 12115; \citealt{paxton2011,2013ApJS..208....4P,2015ApJS..220...15P,2018ApJS..234...34P,2019ApJS..243...10P}). The initial metallicity is Z = 0.034 derived from the observed [Fe/H], assuming the solar metallicity is $Z=0.014$ \citep{2009ARA&A..47..481A}. The detailed model for \hoststar has an initial mass of 1.325 $M_{\odot}$. Combining the initial metallicity, those initial settings yield the sequence of models of the star that are characterized in the $\log\,(g)$ - $T_{\rm eff}$ plane in the bottom panel of Figure \ref{fig:planettides}.

% Figure \ref{fig:sunupdate} illustrates the semimajor axis of currently known population of close-in planets against log($g$) of their host star. Planets confirmed around evolved stars are show to the right of the plot. Black stars correspond to planets discovered before this work, and red stars correspond to planets confirmed by this work. Lines indicate critical semimajor axis values, beyond which dynamical tides overwhelm equilibrium tides and runaway planetary inspiral begins \citep{sun2018}. Both Kepler-91b and \planet are clearly past this critical semimajor axis value, and \planetthree may have passed it as well. 

In the top panel of Figure \ref{fig:planettides}, we show the orbital semimajor axis as a function of log($g$) using the stellar model described in the last paragraph, where log($g$) operates as a proxy for evolutionary state of the host star. The orbital decay rate is calculated by using Equation (1) in \citet{sun2018}. Colored lines illustrate evolutionary pathways for different initial orbital separations, and the red point illustrates the system as seen today, thus indicating that only the light blue model is consistent with the existence of this system. Note that the black solid line corresponds to an initial semimajor axis of 0.04467 AU, which is the current orbital separation of the \hoststar system. At this initial orbital separation, the system should merge at $\log (g) \sim 4$, significantly higher than the observed $\log (g) = 3.5$. This implies \planet began its life at a larger semimajor axis and is undergoing rapid orbital decay as dynamical tides overwhelm equilibrium tides and runaway planetary inspiral occurs \citep{sun2018}. The short orbital period and low surface gravity of the \planet system imply the inspiral process must be ongoing. 

Given the uncertainty in the measurement of the stellar mass and metallicity of the \hoststar system, and the sensitivity of the evolutionary history of the star to the initial mass and metallicity, we determine a range of the rate of change in semimajor axis and orbital period assuming different initial mass and metallicity. Our numerical simulations predict that dynamical tides should dominate over the effect of equilibrium tides, and estimate the orbital decay timescale of the \hoststar system to be between 1 and 10 Myr.

The second panel of Figure \ref{fig:planettides} illustrates the change in tidal energy dissipation as a function of log($g$), where solid lines correspond to equilibrium tides (calculated using Equation (5) from \citet{sun2018}), while dashed lines correspond to dynamical tides (Equation (14) from \citet{sun2018}). Once \hoststar began evolving off the main sequence, dynamical tides dominate the system, resulting in eventual runaway orbital decay. The exact timing of this runaway orbital decay depends both on the orbital properties of the system as well as the internal structure of the host star \citep{weinberg2017}. 

The third panel of Figure \ref{fig:planettides} illustrates stellar rotational frequency (solid lines) and orbital frequency (dashed lines) as a function of log($g$). To simplify the problem, we assume a zero-rotation rate of the star at the zero-age main sequence, but note that this evolution analysis is not sensitive to the initial rotation rate of the star. In the \hoststar system, like the WASP-12 system, the planet \planet should be synchronized. In addition, \planet does not have enough orbital angular momentum to synchronize the host star, and thus the star's spin rate is always smaller than the orbital frequency. Most of the tidal dissipation energy is therefore generated in the non-synchronized primary star. 

Finally, the bottom panel shows the stellar effective temperature and the location of the base of the stellar convection zone as a function of stellar surface gravity. As \hoststar evolves across the subgiant phase, it becomes cooler, its core becomes radiative near log($g$) $\sim$ 4 and its convective envelope grows deeper over time.  The tidal dissipation rate is affected by the exact location of this boundary in the star. The location of this boundary is highly variable during the subgiant and early red giant phase of evolution, thus introducing degeneracies between different models of stars on the subgiant branch \citep{tayar2021}. Future observation of or upper limits on orbital decay in the \hoststar and similar systems will better constrain the value of $Q'_\star$, breaking degeneracies between subgiant stellar models and updating our ability to constrain stellar structure during subgiant and red giant branch stellar evolution.
% Figure \ref{fig:stellinter} illustrates cross-sections of the interior of \hoststar, assuming a literature values for $Q_\star$`. Each column corresponds to the 16th percentile and 84th percentile limits on stellar mass, showing the relative abundances of hydrogen and helium as a function of enclosed mass within the star in the top panels, and the energy released by different fusion processes as a function of enclosed mass in the bottom panels. Different masses imply different core mass fractions and different radiative-convective boundary locations throughout the star. Similarly, 

\section{Conclusions} \label{sec:conclusions}

We have conducted a search for planets transiting evolved stars using \tess\ Full Frame Image data. Our search for planets around evolved stars uncovered \planet, \planettwo, and \planetthree, two of which are the shortest period planets orbiting evolved stars (R$_*$ $>$2 R$_\odot$, T$_\mathrm{eff}$ $<$ 6000 K) found to date. These planets display a diverse range of properties, and offer new glimpses of the final stages of planetary system evolution. Our main conclusions are as follows:

\begin{itemize}
    \item \planet is a massive, uninflated planet (1.6 M$_\mathrm{J}$, 0.9 R$_\mathrm{J}$) on the shortest period orbit (P = 2.9943 d) ever observed around a red giant (3.2 R$_\odot$, 1.4 M$_\odot$) star. Despite its high incident flux, it is not inflated nor on an eccentric orbit, and appears to be inducing ellipsoidal variations in its host star. Finally, based on estimates of tidal inspiral in evolved systems, the tidal decay of this planet should be measurable by \tess\ on a timescale of years. Current constraints indicate orbital decay is not yet measurable in this system, and we constrain the modified tidal quality factor of the star $Q'_\star > 2 \times 10^4$, in agreement with values found in other short-period hot Jupiter systems. Tighter constraints on this value made from $\geq$1 years of observations from the \tess extended mission data may constrain the location of the star's radiative-convective boundary, as it is strongly dependent on the properties of wave propagation through the stellar interior.
    \item \planettwo is a less massive planet (0.5 M$_\mathrm{J}$) on an even shorter orbit (P = 2.923 d) around a subgiant host (2.3 R$_\odot$, 1.5 M$_\odot$). In contrast to \planet, \planettwo is highly inflated (1.5 R$_\mathrm{J}$), and there are no signs of out-of-transit phase variations in the system. Its orbit appears circular. Its high level of inflation and short period make it suitable to atmospheric observation. Transmission spectroscopy of these (and similar) planets will be valuable to test the validity of planet re-inflation and evolution metrics.
    \item \planetthree is the most evolved system in our sample, on the longest period orbit (1.7 R$_\mathrm{J}$, 0.7 M$_\mathrm{J}$ planet orbiting a 4.2 R$_\odot$, 1.3 M$_\odot$ star every 6.2 d). Though the data for this sample is still limited, the brightness of this target also make it potentially amenable for transmission spectroscopy with JWST. In addition, tighter constraint on the eccentricity ($e < 0.15$) observed in the system could place new constraints on planet inflation and migration mechanisms and timescales.
    \item These planets display a wide range of inflation efficiencies. When considered with the previously known population of evolved inflated planets, the evolved planet population appears to be relatively under-inflated for a given incident flux, regardless of planet mass. This indicates that a mass-radius-flux relation may not be sufficient to describe the entire inflated planet population.

\end{itemize}

%    All of these systems feature intermediate-mass stars at various post-main sequence evolutionary states. Continued confirmation and characterization of these and similar planetary systems will reveal the occurrence of planets around an intermediate-mass population of hosts, which has been debated due to the difficulty in identifying these systems. Furthermore, this population will constrain planet re-inflation models, testing the relative effects of tidal and irradiational heating mechanisms with various orbital configurations at different evolutionary states, and planet engulfment models, constraining speeds for planet inflation, inspiral and angular momentum exchange, and thus constraining the location and evolution of the radiative-convective boundary of intermediate-mass stars.

Detailed stellar spectral analysis can reveal evidence for elements that are short-lived at the surfaces in giant stars, such as lithium and refractory elements. Detection of large amounts of lithium in addition to refractory elements with high condensation temperatures such as magnesium, aluminum, calcium and silicon in giant stars is thus an indication of recent planet engulfment \citep{oh2018, soaresfurtado2020}. A more detailed analysis of previously acquired Keck/HIRES spectra of these stars can reveal the relative enhancement of the above elements in these systems. %More RVs from Keck/HIRES will constrain the eccentricity of \planetthree to higher precision. This will further test theories of planet evolution and orbit circularization. Additional radial velocity measurements of all planets in the system could yield more precise mass estimates, which are essential for accurate atmospheric characterizations \citep{batalha2019}.

\tess\ extended mission operations will be essential for improving planet parameters as well as stellar parameters for these systems. This has already been shown to be the case for \planetthree, where \tess\ Extended Mission data both improved the period precision by an order of magnitude and also revealed potential asteroseismic signals of the star. We expect that the higher cadence and longer baseline Extended Mission observations of \planet and \planettwo may allow asteroseismic characterization of these stars as well.

The additional signal and higher cadence available for \tess\ Extended Mission observations will also improve our ability to measure transit parameters, as well as any other out-of-transit phase variability. Such variability has been measured for several main sequence systems with higher cadence data from the \tess\ nominal mission \citep{wong2020,wong2021}.

It has been observed that hot Jupiters orbiting stars cooler than 6250 K tend to have low obliquities \citep{winn2010}. Several hypotheses have been introduced to explain this, such as the damping of inclination by the stellar convective envelope \citep{winn2010} or magnetic realignment of stellar orbits during pre-main sequence evolution \citep{spalding2015}. Given that the systems studied here have evolved from hot stars with thin convective envelopes on the main sequence to giants with thick convective envelopes, measuring the obliquity of these newly confirmed planets can distinguish between these theorized mechanisms for spin-orbit alignment in hot Jupiter systems. Measuring the Rossiter–McLaughlin (RM) effect \citep{mclaughlin1924, rossiter1924} for these systems with extreme precision radial velocity instruments could reveal trends of planetary system obliquities with respect to stellar mass \citep{rubenzahl2021, cabot2021}, and constrain models of hot Jupiter formation \citep[e.g.,][]{dawson2018, albrecht2021}. Stellar obliquity can also be constrained via asteroseismology \citep{huber2013} and thus future asteroseismic detections for these systems, in combination with future RM measurements, could reveal true system obliquities.

%The sky-projected obliquity of these systems can be measured spectroscopically. By measuring the Rossiter–McLaughlin (RM) effect, the anomalous Doppler shift caused by a planet occulting regions with differing projected rotational velocities as it transits the stellar disk \citep{mclaughlin1924, rossiter1924}, we can constrain the planet's sky-projected obliquity. If the planet’s orbit is aligned with the rotation of the star (prograde), its transit will cause an anomalous redshift, followed by an anomalous blueshift. A antialigned (retrograde) orbit will cause the opposite to occur. 

%As the host stars of these systems were likely hotter than 6250 K on the main sequence, the alignment of these systems would suggest that planetary systems can become aligned due to convective damping on timescales comparable to the post-main sequence lifetimes of these systems. 

Transmission spectroscopy of exoplanet atmospheres has been carried out successfully from ground-based facilities \citep{hoeijmakers2020,giacobbe2021}. These observations allowed identification of atmospheric metals and/or molecular species which provide insight into the formation conditions for these planets through elemental and refractory/volatile ratios \citep{lothringer2020}. Observation of emission features before and during planet occultations from ground-based facilities may prove more promising. Indications of atmospheric outflows, first observed using the Hubble Space Telescope \citep{spake2018}, have now been accomplished from ground-based narrow-band photometry \citep{vissapragada2020,paragas2021}. The more extreme environments of these planets warrant a search for atmospheric outflows, which should be occurring at a faster rate and thus may be observable from ground-based facilities. Otherwise, upcoming space-based observatories such as the James Webb Space Telescope (JWST) will likely be required to probe the atmospheric signal of these planets.

%Continued followup of similar planetary candidate systems is crucial to better understanding this population. The dependence of planet inflation and re-inflation on dynamical processes such as orbit circularization and stellar irradiation can be investigated via this population of system. Furthermore, continued \tess\ observations will provide a basis for further constraining the orbital stability of these systems, as well as their eccentricities, through constraint of transit timing and duration variations. Finally, d

Detailed kinematic information for these stars allow investigation into differences between known moving groups and distinct kinematic groups throughout the Milky Way. Given the intrinsic brightness of evolved stars, among magnitude-limited surveys for planets such as \tess, evolved systems will be the first to reveal planetary demographics outside of the thin disk of our Galaxy.

\acknowledgements{

We thank Howard Isaacson and Daniel Foreman-Mackey for helpful discussions. We acknowledge the use of public TESS data from pipelines at the TESS Science Office and at the TESS Science Processing Operations Center. Resources supporting this work were provided by the NASA High-End Computing (HEC) Program through the NASA Advanced Supercomputing (NAS) Division at Ames Research Center for the production of the SPOC data products. This work was supported by a NASA Keck PI Data Award, administered by the NASA Exoplanet Science Institute. Data presented herein were obtained at the W. M. Keck Observatory from telescope time allocated to the National Aeronautics and Space Administration through the agency's scientific partnership with the California Institute of Technology and the University of California. The Observatory was made possible by the generous financial support of the W. M. Keck Foundation. The authors wish to recognize and acknowledge the very significant cultural role and reverence that the summit of Maunakea has always had within the indigenous Hawaiian community. We are most fortunate to have the opportunity to conduct observations from this mountain. S.G., N.S., and D.H. acknowledge support by the National Aeronautics and Space Administration under Grant 80NSSC19K0593 issued through the TESS Guest Investigator Program. D.H. acknowledges support from the Alfred P. Sloan Foundation and the National Aeronautics and Space Administration (80NSSC21K0652), and the National Science Foundation (80NSSC21K0652). N.S., A.C., and M. R. acknowledge support from the National  Science Foundation through the Graduate Research Fellowship Program under Grants 1842402 and DGE-1752134. Any opinions, findings, and conclusions or recommendations expressed in this material are those of the authors and do not necessarily reflect the views of the National Science Foundation. M. S. acknowledges funding support from NSF ACI-1663696 and AST-1716436. T.D.K. acknowledges support from the 51 Pegasi b fellowship in Planetary Astronomy sponsored by the Heising-Simons Foundation. P. D. is supported by a National Science Foundation (NSF) Astronomy and Astrophysics Postdoctoral Fellowship under award AST-1903811. This research has made use of the Exoplanet Follow-up Observation Program website, which is operated by the California Institute of Technology, under contract with the National Aeronautics and Space Administration under the Exoplanet Exploration Program. Funding for the TESS mission is provided by NASA's Science Mission Directorate. 
}

\software{This work relied heavily on open source software tools, and we would like to thank the developers for their contributions to the astronomy community. For data access and de-trending, this research made use of \lightkurve, a Python package for \kepler and \tess data analysis \citep{lightkurve}, \tesscut, a MAST tool for extracting observations from \tess FFIs \citep{brasseur2019}, and \eleanor, a pipeline for producing and de-trending \tess FFI light curves \citep{feinstein2019}. The analysis portion of this research relied on \astropy \citep{astropy2013,astropy2018}, as well as \exoplanet \citep{exoplanet:exoplanet} and its dependencies \citep{exoplanet:agol19, exoplanet:exoplanet, exoplanet:kipping13, exoplanet:luger18, exoplanet:pymc3, exoplanet:theano}.}

%\pagebreak
\nocite{tange2018}
\clearpage
\bibliography{references}{}
\bibliographystyle{aasjournal}

\end{document}